\documentclass[prl,twocolumn,superscriptaddress,secnumarabic,balancelastpage,amsmath,amssymb]{revtex4-1}
\usepackage{amssymb}
\usepackage{amsmath}
\usepackage[dvipsnames]{xcolor}
\usepackage{chngpage}
\usepackage{lgrind}        
\usepackage{color}         
\usepackage{graphics}      
\usepackage[pdftex]{graphicx}      
\usepackage{longtable}     
\usepackage{epsf}          
\usepackage{bm}            
\usepackage{asymptote}     
\usepackage{thumbpdf}
\usepackage{verbatim}
\usepackage{url}
\usepackage{MnSymbol}
\usepackage{physics}
\usepackage[colorlinks=true, hyperindex, breaklinks, linkcolor=blue, urlcolor=blue, citecolor=blue]{hyperref} 
\usepackage{enumitem}	

\usepackage{float}
\usepackage{tipa, extraipa}
\usepackage{times}
\usepackage{color}
\usepackage{verbatim}
\usepackage{soul,xcolor}
\setstcolor{red}
\usepackage{colortbl}
\usepackage{amsthm}
\usepackage{stmaryrd}
\usepackage[linesnumbered, ruled, vlined]{algorithm2e}
\usepackage{makecell}

\newcommand{\dsl}[0]{\llbracket}
\newcommand{\dsr}[0]{\rrbracket}

\usepackage{tabularx}

\usepackage[normalem]{ulem}

\newcommand{\outline}[1]{\section{#1}}

\makeatletter
\AtBeginDocument{
\g@addto@macro{\appendix}{\renewcommand{\p@subsection}{\@Alph\c@section}}
}
\makeatother

\begin{document}

\title{Constant-Overhead Fault-Tolerant Quantum Computation with Reconfigurable Atom Arrays}

\author{Qian Xu}
\thanks{These authors contributed equally.}
\affiliation{Pritzker School of Molecular Engineering, The University of Chicago, Chicago 60637, USA}

\author{J. Pablo Bonilla Ataides}
\thanks{These authors contributed equally.}
\affiliation{Department of Physics, Harvard University, Cambridge, Massachusetts 02138, USA}

\author{Christopher A. Pattison}
\affiliation{Institute for Quantum Information and Matter, California Institute of Technology, Pasadena, CA 91125}

\author{Nithin Raveendran}
\affiliation{Department of Electrical and Computer Engineering, University of Arizona, Tucson, AZ 85721, USA}

\author{Dolev Bluvstein}
\affiliation{Department of Physics, Harvard University, Cambridge, Massachusetts 02138, USA}

\author{Jonathan Wurtz}
\affiliation{QuEra Computing Inc., 1284 Soldiers Field Road, Boston, MA, 02135, US}

\author{Bane Vasić}
\affiliation{Department of Electrical and Computer Engineering, University of Arizona, Tucson, AZ 85721, USA}

\author{Mikhail D. Lukin}
\affiliation{Department of Physics, Harvard University, Cambridge, Massachusetts 02138, USA}

\author{Liang Jiang}
\email{liang.jiang@uchicago.edu}
\affiliation{Pritzker School of Molecular Engineering, The University of Chicago, Chicago 60637, USA}

\author{Hengyun Zhou}
\email{hyzhou@quera.com}
\affiliation{Department of Physics, Harvard University, Cambridge, Massachusetts 02138, USA}
\affiliation{QuEra Computing Inc., 1284 Soldiers Field Road, Boston, MA, 02135, US}

\begin{abstract}
Quantum low-density parity-check (qLDPC) codes can achieve high encoding rates and good code distance scaling, providing a promising route to low-overhead fault-tolerant quantum computing.
However, the long-range connectivity required to implement such codes makes their physical realization challenging.
Here, we propose a hardware-efficient scheme to perform fault-tolerant quantum computation with high-rate qLDPC codes on reconfigurable atom arrays, directly compatible with recently demonstrated experimental capabilities.
Our approach utilizes the product structure inherent in many qLDPC codes to implement the non-local syndrome extraction circuit via atom rearrangement, 
resulting in effectively constant overhead in practically relevant regimes.
We prove the fault tolerance of these protocols, perform circuit-level simulations of memory and logical operations with these codes, and find that our qLDPC-based architecture starts to outperform the surface code with as few as several hundred physical qubits at a realistic physical error rate of $10^{-3}$.
We further find that less than 3000 physical qubits are sufficient to obtain over an order of magnitude qubit savings compared to the surface code, and quantum algorithms involving thousands of logical qubits can be performed using less than $10^5$ physical qubits.
Our work paves the way for explorations of low-overhead quantum computing with qLDPC codes at a practical scale, based on current experimental technologies.
\end{abstract}
\maketitle

\outline{Introduction}
Quantum error correction (QEC) is believed to be essential for realizing large-scale fault-tolerant quantum information processing.
However, traditional schemes for achieving quantum error correction, such as the paradigmatic surface code, are generally very costly in terms of resource overhead, requiring millions of qubits to solve problems of interest~\cite{fowler2012surface,litinski2019game,beverland2022assessing,gidney2019how}.

Recently, a new approach based on high-rate quantum low-density parity-check (qLDPC) codes has been proposed as a promising route to reduce the resources required.
Unlike planar surface codes~\cite{kitaev2003fault, fowler2012surface,litinski2019game} that encode a single logical qubit per block, qLDPC codes can encode multiple logical qubits per block and achieve a much higher, asymptotically constant encoding rate~\cite{tillich2014quantum, panteleev2019degenerate} as well as better distance scaling~\cite{panteleev2022quantum, breuckmann2020balanced, panteleev2022asymptotically}.
However, in order to realize these appealing features, qLDPC codes require long-range connectivity between qubits, making their physical realization challenging~\cite{bravyi2010tradeoffs,baspin2021connectivity,baspin2022quantifying}.
While several proposals have been made for physical implementation of qLDPC codes in superconducting qubit architectures, the required long-range and multi-layer connectivity is considerably beyond both current and medium-term hardware capabilities~\cite{tremblay2022constant,delfosse2021bounds,strikis2022quantum}.

In bringing qLDPC codes into practical use for full-fledged quantum computation, further challenges arise.
A rigorous analysis of the circuit-level fault tolerance of qLDPC codes is lacking, despite some promising numerical evidence~\cite{delfosse2021bounds, tremblay2022constant}.
Also, it is currently unclear if finite-size qLDPC codes can outperform surface codes in near- or medium-term devices with $\lesssim$$10000$ qubits and realistic physical error rates above $10^{-3}$.
Since Gottesman's seminal results demonstrating that qLDPC codes can enable fault-tolerant quantum computing with constant space overhead~\cite{gottesman2013fault}, 
several practical gate constructions have recently been proposed~\cite{cohen2022low, krishna2021fault, breuckmann2022fold, quintavalle2022partitioning}.
However, no studies of the circuit-level performance of these protocols have been carried out to date.
In particular, it is not clear if the performance and low overhead of the qLDPC codes can be maintained during computation in a full circuit-level fault-tolerant setting.

\begin{figure}
\centering
\includegraphics[width=0.5\textwidth]{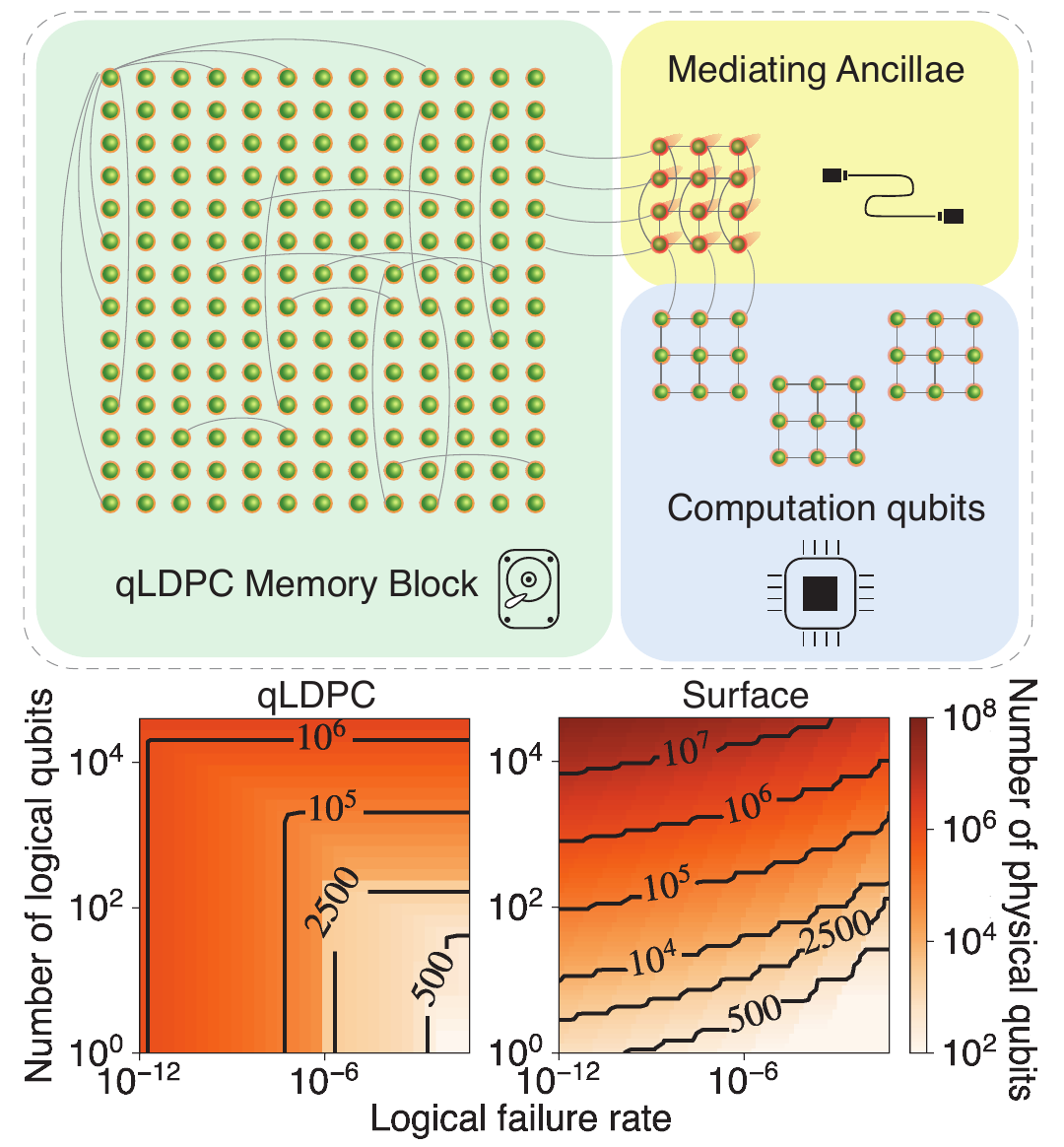}
\caption{\textbf{Architecture of a qLDPC-based fault-tolerant quantum computer using reconfigurable atom arrays.} This consists of a qLDPC memory block, a processor with computational logical qubits, and mediating ancillae between the memory and the processor. The lower panel shows a contour plot of the number of physical qubits (including data and ancilla qubits) required by our architecture, at a $10^{-3}$ physical error rate, given a target number of logical qubits and a target logical failure rate, compared to the surface code. The qLDPC space overhead is given by the minimum of that for the LP codes shown in Fig.~\ref{fig:fig3}(b) with less than $1428$ data qubits and that for HGP codes using an extrapolation of the numerical results in Fig.~\ref{fig:fig3}(a).
    }
\label{fig:architecture}
\end{figure}

In this Article, we propose and analyze a realistic hardware-efficient neutral atom implementation of fault-tolerant quantum computation with high-rate qLDPC codes. We provide concrete experimental and theoretical blueprints, demonstrating their advantage over surface codes starting from as few as several hundred physical qubits.
Our proposal is based on reconfigurable atom arrays, a newly-developed hardware architecture for quantum computation with long-range, reconfigurable connectivity~\cite{bluvstein2022quantum}.
We show how the product structure of many qLDPC codes~\cite{tillich2014quantum,panteleev2022quantum,breuckmann2021quantum} naturally matches the parallelism afforded by physical realizations of reconfigurable atom arrays, enabling their hardware-efficient implementation in a logarithmic number of steps.
Through a combination of single-shot circuit-level threshold proofs and circuit-level simulations, we find competitive performance for hypergraph product (HGP)~\cite{tillich2014quantum} codes and quasi-cyclic lifted product (LP)~\cite{panteleev2022quantum} codes, achieving error thresholds of around 0.6$\%$ under a circuit-level depolarizing noise model that neglects idling errors.
Accounting for idling errors, which only have a minor contribution for the finite-size codes of our interest, we achieve an order of magnitude saving over the surface code with less than 3000 physical qubits (including ancillas) at a physical error rate of $10^{-3}$ (see Fig.~\ref{fig:architecture} bottom panel and Table~\ref{tab:resource_estimate}).
We further extend this analysis to logical gate operation, numerically demonstrating that the high thresholds and good subthreshold scalingof high-rate qLDPC codes can be maintained during computation, paving the way to low-overhead fault-tolerant quantum computing.

\outline{Overview of qLDPC-based Quantum Computer}
An overview of our qLDPC-based approach to fault-tolerant quantum computation is shown in Fig.~\ref{fig:architecture}.
It consists of a high-rate qLDPC memory block that reliably and efficiently stores the quantum information, a processor with computational logical qubits such as surface or color codes that perform logical gates, and mediating ancillae that interconnect the memory and processor. This allows us to take advantage of the dense storage capabilities of the qLDPC block while allowing flexible execution of quantum circuits.
Adopting the conventional $\dsl n,k,d\dsr$ notation for a code with $n$ physical qubits, $k$ logical qubits, and distance $d$, the qLDPC block using the HGP codes described below can provide a dense $\dsl \Theta(k), k, d_{\mathrm{mem}}\dsr$ encoding, where the memory distance  $d_{\mathrm{mem}}=\Theta(\sqrt{k})$.
This results in a constant encoding rate $k/n$ and a logical failure rate (LFR), defined as the probability that any of the logical qubits fails per code cycle, exponentially decaying with the code distance.
We note that using LP codes with a higher encoding rate and better distance scaling (see Methods and Fig.~\ref{fig:fig3}) further reduces the space overhead at small sizes.
Our processor consists of $m$ computational qubits of code parameters $\dsl \Theta(d_{\mathrm{comp}}^2), 1, d_{\mathrm{comp}}\dsr$, where the code distance $d_{\mathrm{comp}}=\Theta(\mathrm{polylog}(kT))$, with $T$ being the depth of the logical circuit to execute, is chosen to produce a sufficiently low error rate.
The mediating ancillae (see Fig.~\ref{fig:teleportation}) have code parameters $\dsl O(d_{\mathrm{comp}}d_{\mathrm{mem}}), 1, \min(d_{\mathrm{comp}},d_{\mathrm{mem}})\dsr$.
By performing ancilla-assisted lattice surgery, the stored logical information can be teleported between any given pair of memory and computation qubits.
Within this architecture, logical gates can be applied in parallel to a subset $m$ of the stored memory qubits in each logical circuit step.
By choosing $m=o(\sqrt{k})$, i.e. $m$ grows slower than $\sqrt{k}$, the physical qubit overhead is dominated by the memory block, achieving a constant encoding rate for quantum computation. 

\outline{Implementation in Neutral Atom Arrays}
We now describe  the hardware-efficient implementation of this qLDPC-based architecture in the atom array platform. Here, qubits are encoded in long-lived spin degrees of freedom of an atom, with seconds-range coherence times~\cite{jenkins2022ytterbium,ma2022universal,barnes2022assembly,bluvstein2022quantum} and high-fidelity single- and two-qubit control ~\cite{evered2023high,ma2023high,scholl2023erasure,jenkins2022ytterbium,ma2022universal,bluvstein2022quantum}.
By shuttling atoms around in optical tweezers, one can reconfigure the processor connectivity during quantum evolution with minimal decoherence~\cite{bluvstein2022quantum,beugnon2007two} and realize parallel two-qubit gate operations with qubits across the whole system.
The coherent shuttling approach features a high degree of parallelism, inherent to multiplexing with optical tools. A particularly powerful tool is the so-called acousto-optic deflectors (AODs), which can simultaneously control the position of rectangular grids composed of thousands of atoms, simply with two control waveforms for the X and Y coordinates~\cite{bluvstein2022quantum}.
These AOD tools are a key enabling technology for qubit transport in atom arrays, and support an inherent ``product structure" that, as we will now show, is well-suited to the implementation of HGP and LP codes.

\begin{figure}
\centering
\includegraphics[width=0.5\textwidth]{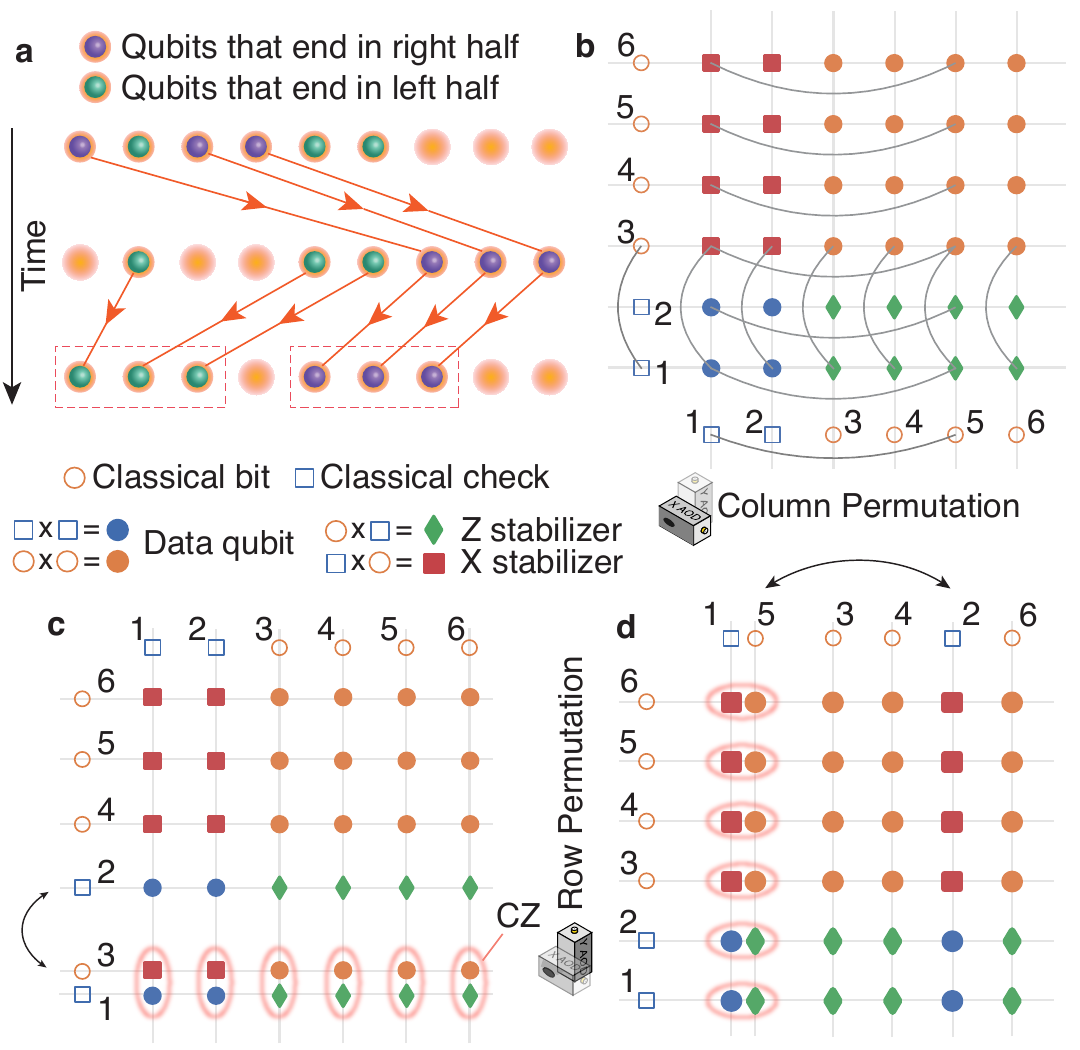}
\caption{\textbf{Efficient implementation of quantum LDPC codes with atom arrays.} (a) Illustration of the algorithm to perform an arbitrary log-depth rearrangement. We first move all atoms that need to end in the right half of the system to the right side, then compact each half into adjacent sites, so that there is sufficient workspace for subsequent steps.
The same procedure can then be repeated on each half of the system recursively for depth $\log(L)$, where $L$ is the length of the atom array to be rearranged, resulting in the desired ordering.
(b) Illustration of the HGP code, obtained as a product of two classical codes. Lines indicate that the parity check at the syndrome node involves the corresponding data node. (c,d) The required connectivity can be implemented via parallel row permutations, followed by parallel column permutations. Although we illustrate this for one pair of row/column interacting, the same permutations and CZs can be applied on multiple rows or columns in parallel.
}    \label{fig:implementation}
\end{figure}

We first describe an algorithm to implement good classical LDPC codes via efficient 1D atom rearrangements without atom collisions.
Recall that a classical (LDPC) code is associated with a parity check matrix $H$, whose rows(columns) are associated with classical checks(bits). The $i$-th check is connected to the $j$-th bit if $H_{i,j} = 1$.
We first lay out the classical LDPC code on a line and group syndrome extraction into layers of parallel entangling operations, resulting in an ordering of checks and bits such that the connected checks and bits in a given layer are neighboring.
To achieve this ordering, we employ a divide-and-conquer rearrangement strategy that only requires a number of steps logarithmic in the total number of checks and bits, as described in detail in Fig.~\ref{fig:implementation}(a) and Methods.
This generalizes prior proposals for scrambling circuits~\cite{hashizume2021deterministic} to arbitrary rearrangements.
After sorting the atoms, a global laser pulse is applied to entangle neighboring checks and bits in parallel, before proceeding to the next layer of atom rearrangement, thus allowing the implementation of syndrome extraction in classical LDPC codes.

We next demonstrate how the product structure of one of the prototypical qLDPC codes (Fig.~\ref{fig:implementation}(b)), HGP codes, naturally matches the product structure of crossed AOD optical hardware, enabling its hardware-efficient implementation.
We start from a pair of classical LDPC codes illustrated in the horizontal and vertical directions, with checks and bits denoted as blue squares and orange circles, respectively.
To construct the resulting HGP code, we place a data qubit at each intersection of a horizontal check and a vertical check, or a horizontal bit and a vertical bit within a 2D grid~\cite{tillich2014quantum} (Fig.~\ref{fig:implementation}(b)).
$X$ stabilizer syndrome qubits and $Z$ stabilizer syndrome qubits are then placed at the intersection of a horizontal check and a vertical bit, or a vertical check and a horizontal bit, respectively.
Importantly, the qubit connectivities are directly inherited from the underlying classical code in the horizontal and vertical directions, and thus the same entangling gates are applied across every row or column, matching well with the product structure of crossed AODs.
Thus, by performing parallel row reordering in the vertical direction based on the vertical LDPC code, interleaved with entangling gates between data and stabilizer qubits (Fig.~\ref{fig:implementation}(c)), and then repeating the same in the horizontal direction (Fig.~\ref{fig:implementation}(d)), we can implement the syndrome extraction for HGP codes. The concrete syndrome extraction circuits are presented in Algs.~\ref{alg:1D}-\ref{alg:pipeline} in Methods.

We now estimate the scaling and quantitative experimental timescales of our rearrangement algorithm, demonstrating that the proposed hardware implementation is indeed achievable with existing experimental parameters.
In Methods, we show that the total rearrangement time scales as $O(\sqrt{L})=O(\sqrt[4]{n})$, where $L$ is the length of the 2D atom array, similar to the scaling for a constant acceleration trajectory.  We estimate that for a moderately-sized HGP code with 10000 qubits, each rearrangement layer between gates requires 3 ms, a small fraction of the coherence time $>10$ s that has been demonstrated in neutral atom arrays~\cite{jenkins2022ytterbium,ma2022universal,barnes2022assembly}.
These timescales can be significantly improved through innovations in optical technologies and compilation. For very large codes, where idling errors are no longer negligible, concatenation with another code can also be employed to extend the effective coherence time~\cite{pattison2023hierarchical}.
Note that HGP codes based on expanding classical LDPC codes (also called quantum expander codes) have the single-shot property, and therefore only a single round of syndrome extraction is required to be fault-tolerant~\cite{leverrier2015quantum,quintavalle2020single,bombin2015single}.
Although we have focused on the implementation of HGP codes, other families of qLDPC codes, such as lifted product (LP) codes, can also be implemented by adapting similar ideas, see Methods for details.

\begin{figure*}[hbt!]
\centering
\includegraphics[width=1\textwidth]{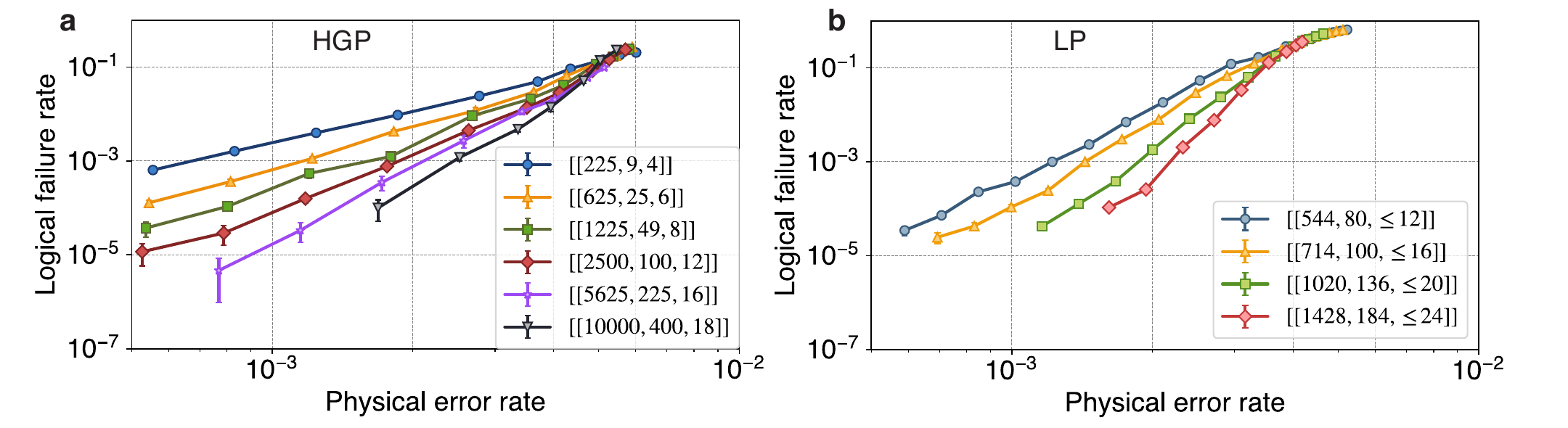}
\caption{\textbf{qLDPC memory performance.} Logical failure rates as a function of physical error rate for the qLDPC memory using HGP codes (a) and LP codes (b), for a depolarizing error model that includes idling errors that increase with the code sizes.
For the HGP codes, we take the hypergraph product of classical codes associated with random $(3,4)$-regular bipartite graphs that have good expanding properties, which have an encoding rate lower bounded by $1/25$.
For the LP codes, we choose $3$ by $5$ base matrices over a quotient polynomial ring and obtain a family of codes with sizes up to $1428$ by increasing the lift size and optimizing the matrix entries. These LP codes have  an encoding rate lower bounded by $2/17$, maintaining a higher encoding rate as well as better distances than HGP codes of the same sizes. See Methods for details of the code construction.
We use a product coloration circuit for syndrome extraction (Alg.~\ref{alg:2D}) and a space-time circuit-level decoder based on BP+OSD that decodes over every $3$ code cycles, regardless of the code sizes.
The LFRs are calculated using $\mathrm{LFR} = 1 - (1- p_L)^{1/m_c}$, where $p_L$ is the total logical failure probability over $m_c$ code cycles. We choose $m_c = 42$ for physical error rates below $4\times 10^{-3}$ and $m_c = 12$ for physical error rates above $4\times 10^{-3}$. $p_L$ is obtained from the Monte Carlo simulations, with a standard deviation $\sqrt{p_L(1 - p_L)/M}$, where $M$ denotes the number of samples.
}
\label{fig:fig3}
\end{figure*}

\setlength{\tabcolsep}{10pt} 
\renewcommand{\arraystretch}{1.2} 
\begin{table*}
\centering
\begin{tabular}{ccccc}
\hline \hline 
\makecell{Logical qubits} & $25$ & $80$ & $180$ & $400$ \\
\hline 
\makecell{Logical failure rate} & $10^{-3}$ & $10^{-4}$ & $2\times 10^{-5}$ & $6\times 10^{-6}$ \\
\hline
\makecell{HGP code physical qubits \\ (improvement over surface code)}
 & $1235\ (1\times)$ & $4606\ (2.8\times)$ & $10760\ (4.0\times)$ & $19600\ (6.9\times)$\\
 \hline
\makecell{LP code physical qubits \\ (improvement over surface code)} & $851\ (1.4\times)$ & $1367\ (9.4\times)$ & $2670\ (16.2\times)$ & \\
\hline 
\hline
\end{tabular}
\caption{Total number of data and ancilla qubits required to reach target numbers of logical qubits and logical failure rates using HGP codes and LP codes, compared to using surface codes. The physical error rate is set to be $10^{-3}$. The estimates for the HGP and LP codes are based on the numerical data in Fig.~\ref{fig:fig3}.}
\label{tab:resource_estimate}
\end{table*}

\outline{qLDPC Memory}
We now analyze the fault-tolerant implementation of HGP and LP codes as a robust quantum memory. In the Supplement~\cite{SM}, we prove the existence of a circuit-level single-shot threshold for qLDPC codes with the linear confinement property~\cite{quintavalle2020single}, under a single-ancilla syndrome extraction circuit and a depolarizing noise model where error rates do not scale with instance size.
The linear confinement property, which requires that for  sufficiently small Pauli errors, the weight of the syndrome increases linearly with the (reduced) weight of the errors, holds for various qLDPC codes decodable by the small-set-flip-type decoders, including HGP codes with sufficiently expanding classical codes~\cite{SM}.

We next supplement this theoretical understanding with numerical simulations of HGP and LP codes at practically-relevant instance sizes~\cite{gidney2021stim}, where we find competitive thresholds and LFRs for both codes.
The details of the code constructions are shown in Fig.~\ref{fig:implementation} and Methods.
We use the product coloration circuit (Alg.~\ref{alg:2D}) for syndrome extraction, a variation of the coloration circuit~\cite{tremblay2022constant} that is more compatible with the product structure of the codes and hardware. The circuit uses a single ancilla for each stabilizer generator and has entangling-gate depth $16$ and $20$ for the HGP and  LP codes we consider.

We construct a space-time circuit-level decoder based on the belief propagation and ordered statistics decoding (BP+OSD) algorithm~\cite{panteleev2019degenerate,roffe2020decoding,higgott2023improved,kuo2022exploiting}.
Specifically, for a QEC circuit with multiple cycles, we construct a bipartite decoding graph~\cite{higgott2023improved,higgott2023improved1,delfosse2023spacetime} over a certain number of cycles, where the check nodes and variable nodes are associated with parities of stabilizer measurement outcomes and circuit faults, respectively.
We apply BP decoding on this decoding graph to infer the circuit fault locations in all noisy code cycles, and apply the BP+OSD decoder in the final round to project back into the code space.
For all memory simulations, we use space-time decoding graphs over three cycles, irrespective of the code size.
Crucially, compared to prior phenomenological decoders that used a simpler decoding graph involving only independent data and measurement errors as bits and decoded over only one code cycle~\cite{grospellier2021combining,delfosse2021bounds}, our space-time circuit-level decoder takes the full circuit details into account and can perform joint decoding on multiple  QEC cycles, improving the threshold~\cite{SM}.
Moreover, the space-time decoding is also crucial for our simulations of logical operations in the next section, where repetitions of syndrome measurements are required for fault tolerance. See Methods for details of the decoder.

In the Supplement~\cite{SM}, we find that the HGP and LP codes have a threshold of $0.63\%$ and $0.62\%$ respectively, under a depolarizing error model without idling errors.
In the sub-threshold regime, assuming the absence of a decoding error floor, the LFRs of the two codes are well approximated by~\cite{SM}
\begin{equation}
\begin{aligned}
    \mathrm{LFR (HGP)} & = 0.07(p_g/0.006)^{0.47n^{0.27}}, \\
    \mathrm{LFR (LP)} & = 2.3(p_g/0.0066)^{0.11n^{0.60}},
    \label{eq:subthreshold_scaling}
\end{aligned}
\end{equation}
indicating that finite-size LP codes have better subthreshold scaling than HGP codes.
When also considering idling errors $p_i(n)$ associated with the atom rearrangement time overhead, which grow as $O(n^{1/4})$ for HGP codes and $O(n^{1/2})$ for LP codes (see Methods for details of the idling error model and the expression for $p_i(n)$ in Eq.~\eqref{eq:idling_error_expression}), there is no asymptotic code threshold~\cite{delfosse2021bounds}.
However, we numerically observe that the effect of adding the idling errors can be approximated by rescaling the gate error $p_g \rightarrow p_g + 3p_i(n)$ using the product coloration circuit (see Supplement~\cite{SM}).
The idling errors have a negligible contribution when $3p_i(n) \ll p_g$, which is the case for current experimental parameters and practically relevant code sizes (see Methods).
Therefore, although there is no asymptotic threshold, constant overhead and fault tolerance can still be achieved at physically-relevant sizes by utilizing the quasi-nonlocal connectivity in atom arrays.
In Fig.~\ref{fig:fig3}(a) and (b), we show the simulated LFR versus the bare two-qubit gate error rate $p_g$ (which we refer to as the physical error rate for simplicity) for the HGP and LP codes, with the idling errors included and rescaled together with $p_g$.
We find that good LFRs  and subthreshold scaling are maintained for both codes in the presence of idling errors.

We now use Eq.~\eqref{eq:subthreshold_scaling} to estimate the total number of data and ancilla qubits $N$ needed to reach a target number of logical qubits $k$ and a target LFR, demonstrating significant advantages of HGP and LP codes over the surface code.
We rescale the gate error $p_g$ to approximate the presence of idling errors, and compare to the surface code subthreshold scaling formula
$\mathrm{LFR} (\mathrm{surface}) = 0.03 k (p_g/0.011)^{\lceil\lfloor \sqrt{n/k} \rfloor / 2 \rceil}$
%
from Ref.~\cite{fowler2010surface,wang2011surface}.
In Table~\ref{tab:resource_estimate}, we present such estimates for finite-size HGP and LP codes that  we directly simulated (see Fig.~\ref{fig:fig3}) at a realistic physical error rate of $10^{-3}$. Both HGP and LP codes outperform surface codes with as few as 25 logical qubits. At a moderate scale of less than 200 logical qubits, LP codes with less than $3000$ physical qubits already achieve a qubit saving of over an order of magnitude. In the lower panel of Fig.~\ref{fig:architecture}, we estimate the space overhead of the HGP codes at a larger scale by extrapolation. We find that HGP codes can also achieve a qubit saving of over an order of magnitude at a scale of 1000 logical qubits and $10^5$ physical qubits.

\outline{Logical operations}
We now present a scheme inspired by Ref.~\cite{cohen2022low} for performing fault-tolerant logical operations, and perform the first numerical simulation of logical gate performance on qLDPC codes using the space-time decoder developed above.
We find that the threshold and logical performance remain almost unchanged when performing gates, indicating that the high threshold and low overhead can be maintained for fault-tolerant quantum computation.

Our scheme is illustrated in Fig.~\ref{fig:teleportation}(a).
We teleport the logical information between the qLDPC memory and ancillary topological codes using a measurement-based circuit (Fig.~\ref{fig:teleportation} (b)), where the prescribed logical measurements are implemented using lattice surgery~\cite{horsman2012surface,cohen2022low,breuckmann2017hyperbolic,poulsen2017fault}.
Universal logical operations can then be performed in the topological codes using standard techniques.
Since each topological code patch and teleportation ancilla patch is much smaller than the qLDPC patch, as long as the number of such patches used is $o(\sqrt{k})$, the space overhead from ancilla patches will be sub-leading.
Note that the scheme in Ref.~\cite{cohen2022low} can also be used for teleportation between a qLDPC block and topological codes. In comparison, our scheme reduces the ancilla patch size by half and thus has smaller space overhead.

As an example, we consider the teleportation from a surface code patch to a HGP patch.
As illustrated in Fig.~\ref{fig:teleportation}(a), this is mediated by an additional ancilla logical qubit, formed by a hypergraph product of two classical codes associated with the logical operators of the two code patches, which enables joint logical measurements by merging code blocks.
We perform $\min\{d_{\mathrm{comp}}, d_{\mathrm{mem}}\}$ rounds of syndrome extraction to ensure tolerance against  measurement errors.
We describe the scheme in more detail in Methods, and prove the fault-tolerance of the scheme under data errors in the Supplement~\cite{SM}. 
\begin{figure*}[t]
\centering\includegraphics[width=1\textwidth]{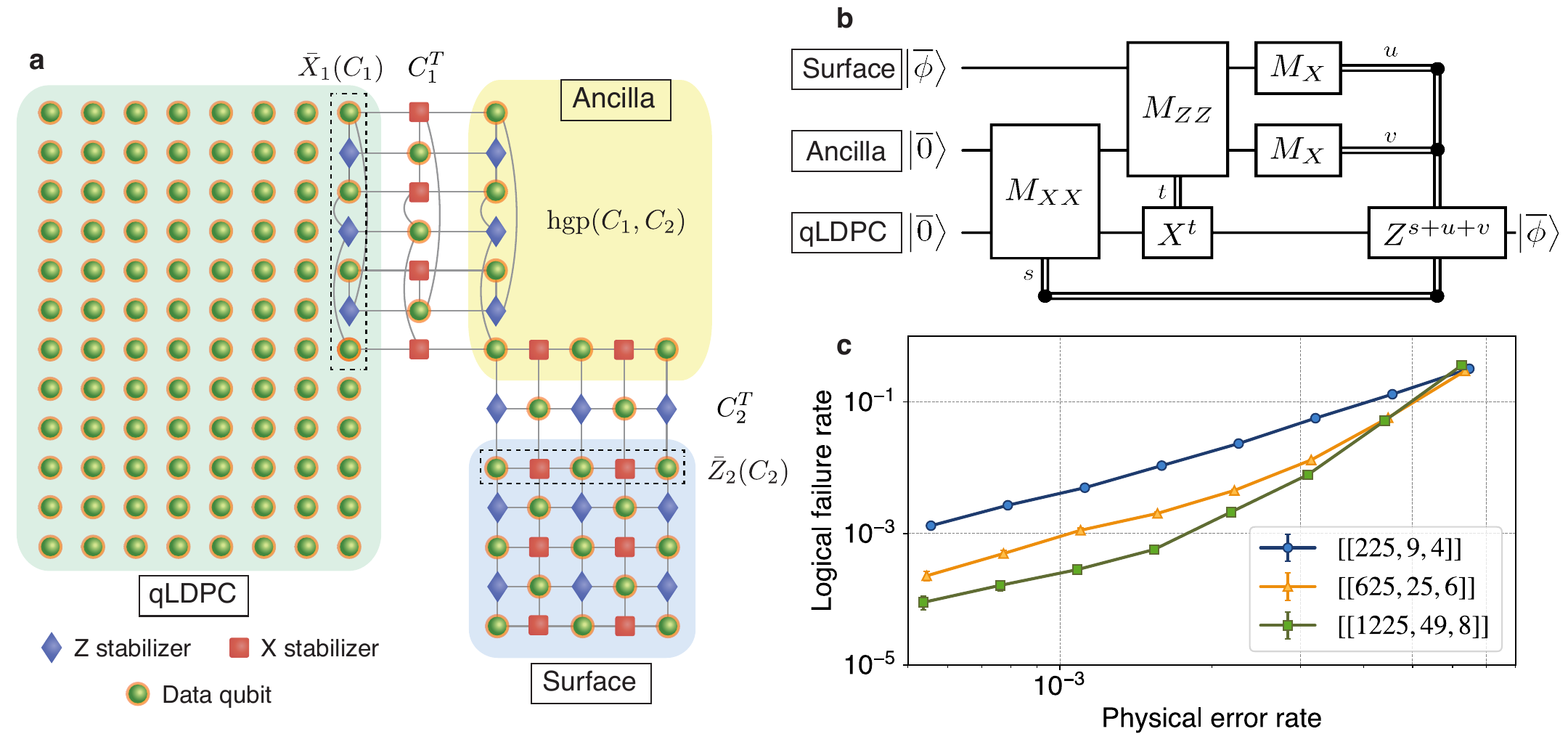}
\caption{\textbf{Fault-tolerant teleportation from surface to qLDPC code.}
We identify the logical $Z$ operator of the surface code, $\overline{Z}_2$, and the logical $X$ operator of one of the qLDPC code's logical qubits, $\overline{X}_1$.
We associate these two logical operators with classical codes $C_1$ and $C_2$,  by mapping the qubits supporting the logical operators to bits and the corresponding incident stabilizer generators to classical checks.
We then construct an ancilla patch as the hypergraph product of $C_1$ and $C_2$, where the columns resemble $C_1$ and the rows resemble $C_2$.
Direct lattice surgery between this ancilla patch and each of the surface and HGP codes is conducted by matching similar boundaries associated with the chosen logical operators. In between similar boundaries, an extra array of ancillary qubits and checks associated with the transpose of the classical code is inserted to mediate the surgery. All checks in the merged code commute with each other, as required, and the product of the stabilizers associated with the checks of the transposed code gives the required joint logical measurement, as in the case of standard surface code lattice surgery~\cite{horsman2012surface}. We elaborate on the lattice surgery procedure in Methods. 
(b) Measurement-based teleportation circuit~\cite{breuckmann2017hyperbolic}. Logical state $\ket{\overline{\phi}}$ is teleported from the surface code to one of the qLDPC's logical qubits. The joint Clifford measurements are conducted through lattice surgery as illustrated on panel (a). (c) Simulated logical failure rates (per code cycle) of the teleportation. Noise is added during the merge and split steps of the $XX$ lattice surgery. We decode with the same space-time circuit-level BP+OSD decoder used in the memory simulations.  The corresponding surface codes paired with the three HGP codes have distances $3$, $5$, and $7$. We record a logical failure if there is an error in any of the logical qubits of the qLDPC code after the teleportation scheme is complete. Denoting the total logical failure probability as $p_L$, we calculate the logical failure rate (per code cycle) as $\mathrm{LFR}=1-(1-p_L)^{2d}$, where there are $2 d$ cycles during the noisy $XX$ lattice surgery, and $d$ denotes the minimal code distance. 
The plotted physical error rates are rescaled to account for idling errors, as explained in Methods.}
\label{fig:teleportation}
\end{figure*}

To validate this method, we perform circuit-level simulations of the above teleportation, enabled by the space-time decoder described in the previous section. In our simulations, we use the teleportation circuit depicted in Fig.~\ref{fig:teleportation}(b) with an initial state of $\ket{\psi} = \ket{0}$, focusing on errors during the merge and split operations in the $XX$ measurement to evaluate the performance of lattice surgery building blocks.
As shown in Fig.~\ref{fig:teleportation}(c), we observe similar LFRs and threshold crossings for the teleportation as for the memory.
Note that each logical qubit will fail at a rate much lower than the LFR of the entire block shown in Fig.~\ref{fig:teleportation}(c), and the system is far below break-even at $10^{-3}$ physical error rate, even for the smallest code of size $225$.
This demonstrates that the high threshold and low resource overhead of the qLDPC code can be maintained at the computation level.

\outline{Discussion and outlook}
By demonstrating large space overhead savings in practical regimes, good performance of logical gate operations, and providing a blueprint for their implementation with existing hardware capabilities, our work brings the use of high-rate qLDPC codes for fault-tolerant quantum computation into the practical regime.

Although our scheme shows a significant reduction in space overhead, it still carries substantial time overhead.
This is because fault tolerance during gate operation requires $\Theta(d)$ QEC cycles, and the low encoding rate of the ancilla and computational code patches limit the logical parallelism when maintaining low space overhead.
We expect certain compilations of quantum algorithms that have limited parallelism to be natural candidates for our architecture~\cite{gidney2019how}.
However, it would be interesting to carry out end-to-end algorithmic compilation with qLDPC codes to evaluate the full space-time cost, and understand which algorithms and compilations are most suited to our architecture.
Another exciting avenue of research is to improve upon the QEC constructions used here, including alternative qLDPC code constructions with better properties, single-shot logical gate constructions~\cite{quintavalle2022partitioning,breuckmann2022fold}, as well as the use of other types of computational logical qubits that may support transversal non-Clifford gates~\cite{bombin2013gauge} or have lower overhead.
Moreover, just as topological codes have interpretations as topological phases of matter, it will be interesting to explore the connection between qLDPC codes and highly entangled states of matter~\cite{anshu2022nlts}, and the techniques described here may be useful for the exploration of novel exotic states of matter.

\outline{Acknowledgements}
We acknowledge helpful discussions with Debayan Bandyopadhyay, Nikolas Breuckmann, Maddie Cain, Larry Cohen, Casey Duckering, Sepehr Ebadi, Simon Evered, Xun Gao, Sasha Geim, Marcin Kalinowski, Sophie Li, Junyu Liu, Tom Manovitz, Balazs Matuz, Nishad Maskara, Quynh Nguyen, Hendrik Poulsen Nautrup, Davide Orsucci, Narayanan Rengaswamy, Michael Vasmer, Han Zheng, among others. We particularly thank Anirudh Krishna for detailed feedback on our results and manuscript.
We are grateful for the support from the University of Chicago Research Computing Center for assistance with numerical simulations.
This work was supported by ARO, ARO MURI, AFOSR MURI, NSF, NTT Research, the Packard Foundation, US Department of Energy (DOE Quantum Systems Accelerator Center), the Center for Ultracold Atoms, the Institute for Quantum Information and Matter (NSF), the DARPA ONISQ program, the NSF Graduate Research Fellowship Program, the Fannie and John Hertz Foundation and the Ramsay Centre for Western Civilisation. The authors declare the following competing interests: M.D.L. is a co-founder and shareholder of QuEra Computing. J.W, H.Z. are employees of
QuEra Computing.

\clearpage
\newpage
\section*{Methods}
\noindent\textbf{Code constructions}\\
We primarily focus on two families of qLDPC codes, which we describe in detail in this section. We leave the extension of these results to other families, such as asymptotically good codes~\cite{panteleev2022quantum, breuckmann2020balanced, panteleev2022asymptotically, leverrier2022quantum, gu2022efficient,dinur2022good, lin2022good}, to future work.

The first family of codes are hypergraph product (HGP) codes \cite{tillich2014quantum}, formed from the product of two classical LDPC codes.
We have provided a geometric sketch of the code properties in the main text, and instead focus here on an alternative, algebraic description of the codes and provide more details of their code properties.
Algebraically, if we denote the parity check matrix (where rows describe bits that should sum to an even number in the absence of errors) of the two underlying classical codes as $H_1\in \mathbb{F}_2^{r_1 \times n_1}$, $H_2\in \mathbb{F}_2^{r_2 \times n_2}$, then the $X$ and $Z$ stabilizer check matrices for the HGP code can be written as
\begin{align}
& H_x= \begin{pmatrix}
H_1^T \otimes I_{r_2} & I_{n_1} \otimes H_2
\end{pmatrix},
\\
& H_z=\begin{pmatrix}
I_{r_1} \otimes H_2^T & H_1 \otimes I_{n_2}
\end{pmatrix}.
\end{align}

For classical $[n_i,k_i,d_i]$ linear codes defined by $r_i = n_i - k_i$ linearly-independent checks ($i=1,2$), the resulting quantum code has parameters $\dsl n_1n_2+r_1r_2, k_1k_2, \min\{d_1,d_2\}\dsr$.
The surface code is a special case of hypergraph product codes, with the classical codes being  repetition codes.
However, by instead choosing classical codes with good vertex expansion , where $k_i=\Theta(n_i)$, $d_i=\Theta(n_i)$, the resulting quantum code (known as quantum expander codes) encodes a linear number of logical qubits $k=\Theta(n)$ and has distance $d=\Theta(\sqrt{n})$ \cite{leverrier2015quantum}.
Such classical expander codes can be obtained asymptotically, for example, from random biregular Tanner graphs, and will have sufficient vertex expansion with high probability \cite{richardson2008modern}.
Logical operators are inherited from the underlying classical code, and one can choose a basis such that each logical operator has support in only a single row or column~\cite{quintavalle2022partitioning,quintavalle2022reshape}.

In this work, we follow the procedure of Ref.~\cite{grospellier2021combining} and construct HGP codes by taking the hypergraph product of classical LDPC codes defined by $(3,4)$-regular Tanner graphs, i.e. bipartite graphs with degree-$3$ bit nodes and degree-$4$ check nodes.
By increasing the size of the graph, we obtain a family of HGP codes with a constant encoding rate $k/n \geq 0.04$. For each code size, we pick the classical code with the largest distance, Tanner graph girth at least $6$ (length of the shortest cycle in the Tanner graph, obtained through rejection sampling without performing edge swaps), and the largest spectral gap (the gap between the largest two singular values of the classical check matrices) from randomly generated instances.
%
%
It is known that the hypergraph product of vertex-expanding classical codes gives HGP codes that satisfy the confinement property, and support single-shot QEC~\cite{quintavalle2020single, leverrier2015quantum,fawzi2017efficient}.

The second family of codes we consider are quasi-cyclic lifted product (LP) codes~\cite{panteleev2019degenerate,breuckmann2020balanced,raveendran2022finite}, which can be viewed as a hypergraph product code followed by a symmetry reduction to reduce the number of required qubits~\cite{breuckmann2021quantum}.
Algebraically, a quasi-cyclic LP code is obtained from two base protographs (analogs of the classical codes in the HGP construction) associated with two base matrices $\mathbf{B}_1$ and $\mathbf{B}_2$ over the quotient polynomial ring $\mathbb{R}[x]/(x^l-1)$~\cite{panteleev2022quantum}.
Suppose the two base matrices are of size $m_{B_1}\times n_{B_1}$ and $m_{B_2}\times n_{B_2}$, respectively. We obtain two matrices (over the same polynomial ring) $\mathbf{B}_x$ and $\mathbf{B}_z$ by taking the hypergraph product: 
\begin{equation}
\begin{aligned}
\mathbf{B}_x & = \begin{pmatrix}
        \mathbf{B}_1^T \otimes \mathbf{I}_{m_{B_2}} & \mathbf{I}_{n_{B_1}} \otimes \mathbf{B}_2
    \end{pmatrix},\\
    \mathbf{B}_z & = \begin{pmatrix}
       \mathbf{I}_{m_{B_1}} \otimes \mathbf{B}_2^T &   \mathbf{B}_1 \otimes  \mathbf{I}_{n_{B_2}}   \\
    \end{pmatrix}.
\end{aligned}
\label{eq:LP_check_matrices}
\end{equation}
The $X$ ($Z$) check matrix $H_x$ ($H_z$) is then obtained by replacing each entry of $\mathbf{B}_x$ ($\mathbf{B}_z$) with its matrix representation as $l$ by $l$ circulant matrices, a process known as a lift.
The code size is $N = l(n_{B_1}n_{B_2}+m_{B_1}m_{B_2})$ and the number of $X$ and $Z$ checks are $M_x = l n_{B_1} m_{B_2}$ and $M_z = l m_{B_1} n_{B_2}$, respectively. 
The encoding rate is lower bounded by $(N - M_x - M_z)/N = (n_{B_1}n_{B_2} + m_{B_1}m_{B_2} - m_{B_1}n_{B_2} - n_{B_1}m_{B_2})/(n_{B_1}n_{B_2} + m_{B_1}m_{B_2})$.

\begin{figure*}
\begin{center}
\includegraphics[width=1\textwidth]{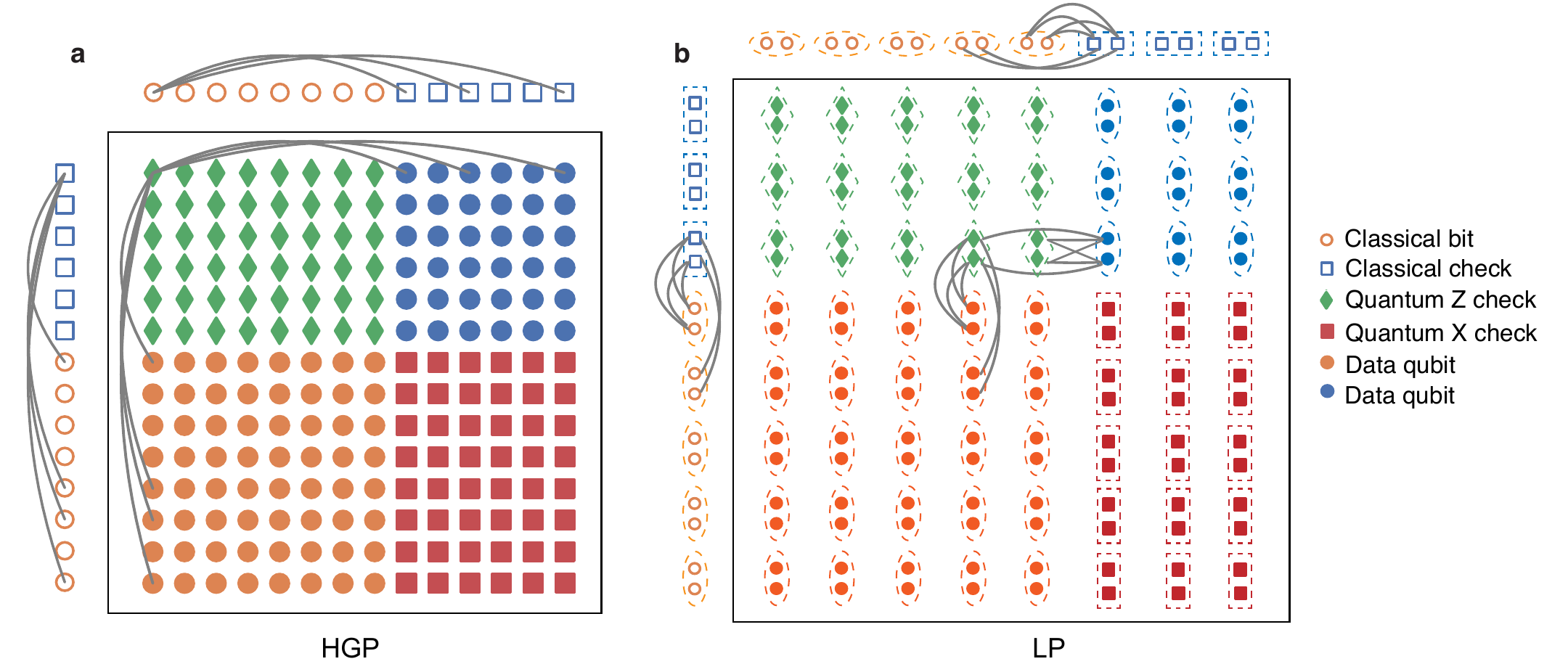}
\caption{{\bf Product structure of HGP codes and LP codes.} (a) The HGP code is constructed from two classical LDPC codes. The classical codes are illustrated on the left and top, where circles indicate classical bits and squares indicate classical checks. A data qubit is placed at each intersection of two classical bits (filled orange circles) and of two classical checks (filled blue circles). $Z$ stabilizer generators are placed at the intersection of horizontal bits and vertical checks, while $X$ stabilizer generators are placed at the intersection of horizontal checks and vertical bits. Each stabilizer is connected to data qubits along the same row or column, with the same connectivity as the classical codes, as illustrated for the top left Z stabilizer. We have omitted other connections for ease of visualization. (b) The LP code is constructed by taking a lift over the hypergraph product of two classical protographs. The protographs and their hypergraph product are indicated by the dashed nodes and the lift is illustrated by the multiple inner nodes within each dashed node. The inner connectivity between two dashed nodes is given by the circulant-matrix representation of the ring elements in Eq.~\eqref{eq:LP_check_matrices}. When flattening the inner nodes vertically (horizontally), the vertical (horizontal) connectivity between the qubits and the checks for each column (row) is the same as the left (top) lifted classical code. }
\label{fig:hypergraph_lifted_product}
\end{center}
\end{figure*}

We can also describe the above construction using graphs. As an example, Fig.~\ref{fig:hypergraph_lifted_product}(b) shows a LP code using a $3$ by $5$ protograph associated with a base matrix $\mathbf{B} \in \{\mathbb{R}[x] /\left(x^2-1\right)\}^{3\times 5}$. The checks and bits of the protograph are illustrated by the big dashed nodes. The $i$-th dashed check node is connected to the $j$-th dashed bit node if $\mathbf{B}_{ij}$ is non-zero. A lift of the protograph is done by replacing each dashed node with its two inner solid nodes, and setting up the connectivity between the inner nodes according to the matrix representation of each ring element $\mathbf{B}_{ij}$. Eq.~\eqref{eq:LP_check_matrices} corresponds to taking the hypergraph product between the protograph and itself, obtaining a grid of dashed nodes. Similar to the hypergraph product code, the connectivity between the dashed nodes (the entries of $\mathbf{B}_x$ and $\mathbf{B}_z$) is inherited from $\mathbf{B}$. Then the qubits and the quantum checks are given by the inner nodes after the lift, and their connectivity is given by the matrix representation of $\mathbf{B}_x$ and $\mathbf{B}_z$. An important feature of the LP codes is that they still have some remaining product structure even after the lift. As shown in Fig.~\ref{fig:hypergraph_lifted_product}(b), when flattening the inner nodes vertically (horizontally), the vertical (horizontal) connectivity between the qubits and the checks for each column (row) is the same as the left (top) lifted classical code.

For the LP codes constructed in this work, we choose a base matrix of dimension 3 by 5, where all entries are monomials, and obtain a family of codes with sizes up to $1428$ by increasing the lift size $l$ from $16$ to $42$.
The classical parity checks are optimized by choosing the base matrix entries over the quotient polynomial ring to obtain the best classical distance for the particular lift size $l$. The choice of the base matrix entries is also such that the girth is at least 8, and the distance of the lifted qLDPC codes matches the designed classical distances with a high probability. Allowing multiple polynomial terms for each base matrix entry and more protographs of different sizes gives more flexibility in LP code design and improves their distances. Such general code constructions and their impact on the proposed scheme will be explored in future work.
Here, we explicitly provide the classical base matrices used to construct the four LP codes used in this work. Denoting $\mathbf{B}^l_{d}$ as a base matrix with a lift size $l$ and a classical code distance $d$ after the lift, the base matrices are

\renewcommand{\arraycolsep}{1.5pt}
\begin{align}
\begin{aligned}
\mathbf{B}^{16}_{12} & =
\begin{bmatrix}
1 &  1  &  1  &  1 & 1 \\
1 &  x^2  &  x^4 &  x^7 & x^{11} \\
1 &  x^3 &  x^{10} &  x^{14} & x^{15}
\end{bmatrix},
\end{aligned}
\begin{aligned}
\mathbf{B}^{21}_{16} & =
\begin{bmatrix}
1 &  1  &  1  &  1 & 1 \\
1 &  x^4  &  x^5 &  x^7 & x^{17} \\
1 &  x^{14} &  x^{18} &  x^{12} & x^{11}
\end{bmatrix},
\end{aligned}\\
\begin{aligned}
\mathbf{B}^{30}_{20} & =
\begin{bmatrix}
1 &  1  &  1  &  1 & 1 \\
1 &  x^2  &  x^{14} &  x^{24} & x^{25} \\
1 &  x^{16} &  x^{11} &  x^{14} & x^{13}
\end{bmatrix},
\end{aligned}
\begin{aligned}
\mathbf{B}^{42}_{24} & =
\begin{bmatrix}
1 &  1 &  1 &  1 & 1 \\
1 &  x^6 & x^7 &  x^9 & x^{30} \\
1 &  x^{40} &  x^{15} &  x^{31} & x^{35}
\end{bmatrix}.
\end{aligned}
\end{align}
These codes have an encoding rate lower bounded by $2/17$ by counting the number of qubits minus the number of checks. For all the resource estimates involving these LP codes, we use $k \approx 0.38n^{0.85}$ that fits well on the above four codes. The quantum code distances are upper bounded by the classical code distances of the above (lifted) base matrices. Through an extensive search for minimum-weight logical operators using a GAP package~\cite{pryadko2022qdistrnd}, we believe these upper bounds are tight.

\noindent\textbf{Atom rearrangement algorithm}\\
The reconfigurable atom array platform features efficient, parallel control and rearrangement of large numbers of qubits, enabling the implementation of long-range connected quantum processors.
As discussed in the main text, optical tools such as crossed acousto-optic deflectors (AODs) can generate a rectangular grid of optical tweezers that can be reconfigured on the fly~\cite{bluvstein2022quantum,beugnon2007two,dordevic2021entanglement}, allowing the control of large code blocks consisting of thousands of physical qubits with only a handful of classical controls.

However, the use of AODs for dynamic rearrangement comes with two key constraints:
First, as the X and Y direction optical spots are controlled by separate AODs, the same operation needs to be applied across multiple rows and/or columns.
Second, different rows of atoms cannot cross each other due to beating between RF tones and atom collisions, although they can be temporarily transferred and stored in static traps, such as those based on spatial light modulators (SLMs).
Thus, the implementation of qLDPC codes with improved code parameters (number of encoded qubits, code distance), which often relies on pseudorandom expander graphs with complex connectivity graphs, requires the development of efficient atom rearrangement algorithms.

We provided a sketch of our atom rearrangement algorithm in the main text, and provide a more detailed description in Algs.~\ref{alg:1D}-\ref{alg:pipeline}.
We also illustrate the algorithm with a movie in Supplemental Movie 1.

\begin{figure*}
\begin{center}
\includegraphics[width=0.5\textwidth]{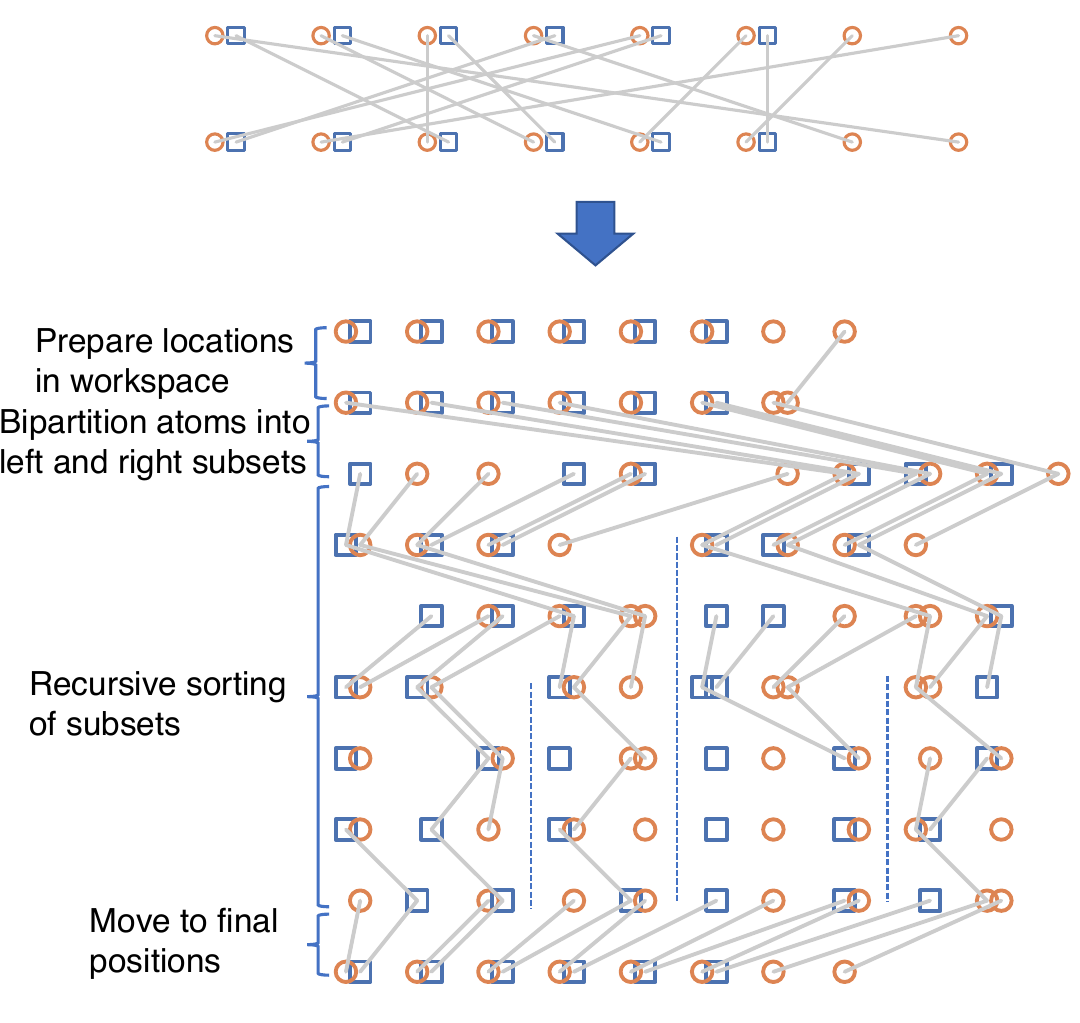}
\caption{{\bf Efficient non-intersecting rearrangement in log-depth.} By using a divide and conquer algorithm, we can perform an arbitrary 1D rearrangement in depth logarithmic in the number of qubits. Repeating this across the array yields an efficient implementation of the desired rearrangements, without requiring intersecting atom trajectories that may lead to additional loss and decoherence. Here, we illustrate the full set of movements required in a small example.}
\label{fig:efficient_rearrangement}
\end{center}
\end{figure*}

The first component, arbitrary 1D atom rearrangements with a number of steps that scales logarithmically, is described in detail in Alg.~\ref{alg:1D}, illustrated in Fig.~\ref{fig:implementation} and explicitly worked out for a small example in Fig.~\ref{fig:efficient_rearrangement}.
Since successive layers each reduce the system size by half, the total number of layers required to achieve the desired rearrangement is $\lceil\log_2 L\rceil$.
Thus, arbitrary rearrangements in very large systems can be achieved in a small number of layers.
We note that although this method shares some similarities to existing techniques such as bitonic sorting, the different constraints (comparators vs. parallel moves) lead to differences in the algorithm itself.
Our algorithm can also be readily extended to use parallel qubit swaps~\cite{litinski2022active,malinowski2023how} at increasing distances as the basic primitive, with the same $O(\log L)$ scaling with system size.

\SetKwInOut{Input}{Input}
\SetKwInOut{Output}{Output}
\SetKwFunction{FMain}{Rearrange}
\SetKwFunction{Fappend}{append}
\SetKwProg{Fn}{Function}{:}{}
\RestyleAlgo{ruled}

\begin{algorithm}
\caption{Arbitrary 1D Atom Rearrangement in a Logarithmic Number of Steps}\label{alg:1D}
\Input{Final ordering $O=[o_i]$ ($i=1...N$) of all $N$ atoms, where $o_i$ is the final position of the atom that was initially at position $i$.}
\Input{Initial positions $A=[a_i]$ ($i=1...N$) of all $N$ atoms, with the positions ordered as $a_1<a_2<...<a_N$.}
\Input{Positions $P=[p_j]$ ($j=1...M$) of all $M$ possible\\ qubit locations, where $M\geq 3N/2$.}
\Output{Positions $C=[c_{s,i}]$ for the $i$th atom in the $s$th rearrangement step, for all $N$ atoms and all rearrangement steps.}
\Fn{\FMain{$O,A,P$}}{
\If{$N=1$}{$c_{1,1} \gets p_1$\\
\Return $[c_{s,i}]$}
$s \gets 1$ \tcp{Layer counter}
\If{$a_N>p_N$}{
\tcp{Compactify atoms to the left to make space for subsequent moves}
\For{$i \gets 1$ \KwTo $N$}{
$c_{s,i} \gets p_i$
}
$s \gets s+1$
}
\tcp{Determine whether each atom ends in the left or right half}
$L, R \gets []$\\
\tcp{Workspace separator for recursion}
$X=\lfloor 3N/2\rfloor$\\
\For{$i \gets 1$ \KwTo $N$}{
\If{$O_i \leq \lfloor N/2\rfloor$}{
L.{append}($i$)\\
$c_{s,i} \gets p_{N+len(L)}$ \tcp{Move to right} 
$c_{s+1,i} \gets X+len(L)$ \tcp{Compactify}
}
\Else{
R.{append}($i$)\\
$c_{s,i} \gets a_i$ 
\tcp{Stay in place}
$c_{s+1,i} \gets len(R)$ \tcp{Compactify}}
}
$s \gets s+2$\\
\tcp{Recursive call on each half}
$C_l \gets$ \FMain{$O[L],C[s-1,L],P[1..X]$}\\
$C_r \gets$ \FMain{$O[R],C[s-1,R],P[X+1..M]$}\\
$C[s..,L] \gets C_l$\\
$C[s..,R] \gets C_r$\\
\Return $C$
}
\tcp{Main function}
\FMain{$O,A,P$}
\end{algorithm}

The second component is the observation that the product structure of crossed AODs matches well with the product structure present in many qLDPC codes.
In addition to the discussion in the main text, we provide the details of the syndrome extraction circuit for HGP codes based on this observation in Alg.~\ref{alg:2D}, which we refer to as the ``product coloration circuit", as it makes use of coloration circuits for each of the component classical codes~\cite{tremblay2022constant}.
Note that the use of our product coloration circuit, as opposed to the coloration or cardinal circuits in Ref.~\cite{tremblay2022constant}, is necessary to fully exploit the parallel rearrangement capabilities across rows and columns.
Here, the native entangling gate set of current atom array systems are diagonal~\cite{levine2019parallel,evered2023high}, so we use CZ gates and appropriate Hadamard rotations to perform syndrome extraction. Note that under global laser excitation and phase advances, any pair of qubits that are within a certain radius (known as the blockade radius) of each other will execute a CZ gate, while any individual qubits will undergo an identity gate. In order to compare our results against the literature, we use CNOT gates as the entangling gates in our simulations. This can be physically justified if the CZ gates are much noisier than the Hadamard gates.

The product coloration circuit separately extracts the $X$ and $Z$ syndromes, each requiring both a horizontal and vertical step.
Thus, if the coloration of each of the classical codes involves $\Delta_C$ colors (for the codes constructed from $(3,4)$-biregular graphs that we considered, $\Delta_C=4$~\cite{delfosse2021bounds}), the product coloration circuit will have $4\Delta_C$ entangling layers. 

The product coloration circuit can also be applied to the LP codes we use in this work. As shown in Fig.~\ref{fig:hypergraph_lifted_product}(b), a LP code has the same product vertical (horizontal) connectivity as a HGP code when flattening the inner nodes vertically (horizontally). Thus, the same product coloration circuit can be applied to the LP codes with an extra step of flattening the inner codes in between establishing the horizontal/vertical connections. As we use $3$ by $5$ base matrices with all weight-one entries, the product coloration circuit for the LP codes has an entangling gate depth of $4\times 5 = 20$.

\begin{figure*}
\begin{center}
\includegraphics[width=0.79\textwidth]{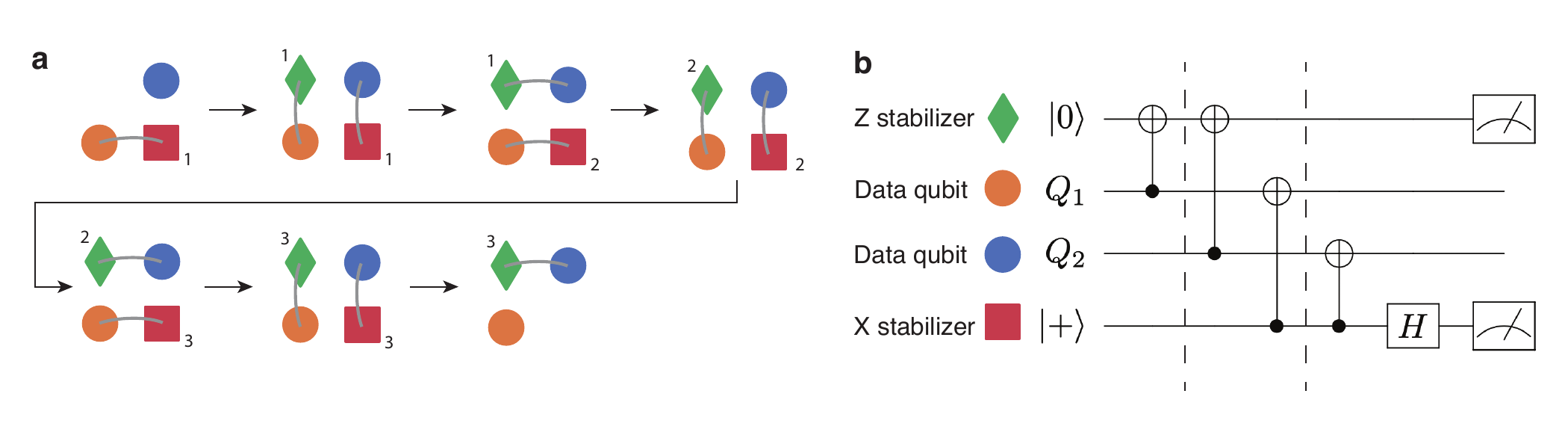}
\caption{\textbf{Illustration of ordering of operations in pipelined syndrome extraction.} (a) Successive steps of entangling gates for the pipelined product coloration circuit described in Alg.~\ref{alg:pipeline}, with $d=3$ rounds of syndrome extraction. Numbers at the corners of the $X$ and $Z$ ancilla qubits denote the round of syndrome extraction they correspond to.
(b) Illustration of a local circuit that data qubits and ancilla qubits  see, with dashed lines indicating different circuit moments. As the $X$ stabilizer interacts with both qubits before the $Z$ stabilizer, the syndrome extraction order is valid. Similar analysis can be performed for the commutation relations with the next round of ancilla qubits.
}
\label{fig:pipeline}
\end{center}
\end{figure*}

To further reduce the depth of the syndrome extraction circuit, we also propose a modification of the above circuit in Alg.~\ref{alg:pipeline} and Fig.~\ref{fig:pipeline}, which we refer to as the pipelined product coloration circuit.
Here, the main challenge is to choose a gate ordering such that the desired $X$ and $Z$ syndromes are correctly extracted.
By performing pipelining and extracting the $X$ syndrome of the second round simultaneously with the $Z$ syndrome of the first round, we can ensure that the gate ordering is always valid, while reducing the number of entangling layers required to perform $d$ rounds of syndrome measurement to $(2d+2)\Delta_C$.
This could be particularly relevant in further suppressing the effect of idling errors as well as improving the performance of logical gates, which in our scheme require $d$ rounds of repetition.
However, our numerical simulations all make use of the product coloration circuit and we leave the exploration of other syndrome extraction circuits to future work.

\begin{algorithm}
\caption{Product Coloration Circuit for HGP Syndrome Extraction}\label{alg:2D}
\Input{Edge colorations $\mathcal{C}_h,\mathcal{C}_v$ of Tanner graphs associated with horizontal and vertical classical codes $C_h$, $C_v$ that form the hypergraph product code.}
\Output{Measurement outcomes of all $X$ and $Z$ stabilizer generators.}
\tcp{$X$ stabilizers}
Apply a Hadamard on all data qubits.\\
Prepare an ancilla in $|+\rangle$ for each $X$ stabilizer and move all $X$ ancilla qubits (red) into the LDPC grid region shown in Fig.~\ref{fig:hypergraph_lifted_product}. Do not include any $Z$ ancilla qubits (green).\\
\For{direction $f \in $ \{horizontal, vertical\}}{
\For{color $c \in \mathcal{C}_f$}{
Apply algorithm 1 in direction $f$, across the whole grid, to bring each pair of qubits connected by an edge of color $c$ in direction $f$ together.\\
Apply a CZ gate on each pair of neighboring qubits.
}}
Apply a Hadamard on all data qubits.\\
Move $X$ ancilla qubits out of the grid region and measure them in the $X$ basis.\\
\tcp{$Z$ stabilizers}
Prepare an ancilla in $|+\rangle$ for each $Z$ stabilizer and move all $Z$ ancilla qubits (green) into the LDPC grid region shown in Fig.~\ref{fig:hypergraph_lifted_product}. Do not include any $X$ ancilla qubits (red).\\
\For{direction $f \in $ \{horizontal, vertical\}}{
\For{color $c \in \mathcal{C}_f$}{
Apply algorithm 1 in direction $f$, across the whole grid, to bring each pair of qubits connected by an edge of color $c$ in direction $f$ together.\\
Apply a CZ gate on each pair of neighboring qubits.
}}
Move $Z$ ancilla qubits out of the grid region and measure them in the $X$ basis.
\end{algorithm}

\begin{algorithm}
\caption{Pipelined Product Coloration Circuit for Multi-round HGP Syndrome Extraction}\label{alg:pipeline}
\Input{Edge colorations $\mathcal{C}_h,\mathcal{C}_v$ of Tanner graphs associated with horizontal and vertical classical codes $C_h$, $C_v$ that form the hypergraph product code.}
\Input{Number of syndrome repetition rounds $d$.}
\Output{Measurement outcomes of all $X$ and $Z$ stabilizer generators.}
\tcp{$X$ stabilizers of the first round}
Apply a Hadamard on data qubits in the bottom left block (orange) of Fig.~\ref{fig:hypergraph_lifted_product}.\\
Prepare an ancilla in $|+\rangle$ for each $X$ stabilizer and move all $X$ ancilla qubits (red) into the LDPC grid region shown in Fig.~\ref{fig:hypergraph_lifted_product}. Do not include any $Z$ ancilla qubits (green).\\
\For{color $c \in \mathcal{C}_h$}{
Apply algorithm 1 in the horizontal direction, across all rows, to bring each pair of qubits connected by an edge of color $c$ in the horizontal direction together.\\
Apply a CZ gate on each pair of neighboring qubits.
}
\tcp{Parallel syndrome extraction for $d-1$ rounds}
\For{$i \gets 1$ \KwTo $d-1$}{
\For{direction $f \in $ \{vertical, horizontal\}}{
Apply a Hadamard on all data qubits.\\
\If{$f == $ vertical}{
Measure any old $Z$ ancillas in the $X$ basis and prepare a fresh ancilla in $|+\rangle$ for each $Z$ stabilizer.\\
}
\Else{Measure any old $X$ ancillas in the $X$ basis and prepare a fresh ancilla in $|+\rangle$ for each $X$ stabilizer.\\}
Move all $X$ and $Z$ ancilla qubits into their appropriate positions in Fig.~\ref{fig:hypergraph_lifted_product}.\\
\For{color $c \in \mathcal{C}_f$}{
Apply algorithm 1 in direction $f$, across all columns, to bring each pair of qubits connected by an edge of color $c$ in direction $f$ together.\\
Apply a CZ gate on each pair of neighboring qubits.
}
}
}
\tcp{Z stabilizers of the final round}
Move $X$ ancilla qubits out of the grid region and measure them in the $X$ basis.\\
\For{color $c \in \mathcal{C}_v$}{
Apply algorithm 1 in vertical direction, across the whole grid, to bring each pair of qubits connected by an edge of color $c$ in vertical direction together.\\
Apply a CZ gate on each pair of neighboring qubits.
}
Move $Z$ ancilla qubits out of the grid region and measure them in the $X$ basis.
\end{algorithm}

\noindent\textbf{Estimation of rearrangement time and effect of idling errors}\\
For the schemes described above, we can estimate the amount of time required to implement one round of stabilizer measurements using the technology that has been demonstrated in Ref.~\cite{bluvstein2022quantum}.
We assume a transfer time $\tau_t$ between a static SLM trap and dynamic AOD trap, a peak atom moving acceleration rate of $a_p$ with a cubic spline trajectory, and a uniform grid spacing $d$.
For simplicity, we assume that the number of atoms on a line to be rearranged is a power of 2, $L=2^k$.
The algorithm will therefore have a depth of $k=\log_2 L$ layers.
In order to provide enough workspace for shuttling, the total number of traps is $3L/2$.

The compactification step at scale $s$ requires a move of distance at most $sd/2$.
Moving all target atoms to the right requires a move of distance at most $sd$.
We can combine the two steps in a way that we pick up atoms for the next move and drop off atoms from the previous move at the same time; thus, each step requires on average one trap transfer between static and dynamic traps.
As described in Ref.~\cite{bluvstein2022quantum}, using a cubic spline movement trajectory, a move of distance $l$ requires time $\sqrt{6l/a_p}$.
The scaling with distance can be understood by the fact that the derivative of acceleration (jerk) causes motional heating, and to maintain a fixed amount of heating in a move, the integrated heating (with dimensions of acceleration) should remain constant, thus giving a time scaling with distance that is the same as a constant acceleration profile.

The total time required for one layer of full rearrangement before each layer of entangling gates in a syndrome extraction circuit is thus
\begin{align}
t_{\mathrm{rearrange}}&=2k\tau_t+\sum_{i=0}^k \qty(\sqrt{\frac{6\cdot 2^id}{a_p}}+\sqrt{\frac{3\cdot 2^id}{a_p}}) \nonumber\\&<2\tau_t\log L+(3+2\sqrt{2})\sqrt{\frac{6Ld}{a_p}}.
\label{eq:rearrangement_time}
\end{align}
Recent experiments have demonstrated parameters on the order of $\tau_t=50\,\mu$s, $a_p=0.02\,\mu$m/$\mu$s$^2$, $d=5\,\mu$m. For a moderately sized code consisting of 10000 qubits (including both data and ancilla qubits), we have $L\approx 100$.
The total trap transfer time is 0.7 ms and the atom movement time is 2.3 ms, for each gate layer.
Assuming a $(3,4)$-biregular graph for the underlying classical expander code, we need 8 rounds of rearrangement to measure one full round of stabilizers for Alg.~\ref{alg:pipeline}, resulting in a total time overhead of 24 ms, a small fraction of the coherence time $T_c > 10$ s that has been demonstrated in neutral atom arrays~\cite{jenkins2022ytterbium,ma2022universal,barnes2022assembly}.
This timescale is somewhat longer than the typical readout timescales, and thus the code cycle time will be dominated by the rearrangement time. The ancilla measurements can be pipelined to happen simultaneously with the atom rearrangements of the following round, and therefore will not increase the run time.

For LP codes, we need to first “flatten” the code layout before implementing the parallel rearrangement scheme. Since we use a fixed $3$ by $5$ protograph, the flattened rectangular array has dimensions roughly of $2n/8$ by $8$. This can be achieved in log depth using a similar divide-and-conquer algorithm that flattens the code by half each time. For example, as shown in Fig.~\ref{fig:hypergraph_lifted_product}(b), we flatten the codes vertically before establishing vertical connections between atoms via row permutations. The vertical connectivity is then the same as an HGP code, and we can use the efficient 1D rearrangement scheme described earlier. Therefore, we estimate the rearrangement time for LP codes by setting $L$ to $n/8$ in Eq.~\eqref{eq:rearrangement_time}.

The rearrangement time in Eq.~\eqref{eq:rearrangement_time} determines the idling errors between sequences of entangling gates in a syndrome extraction circuit. In general, $t_{\mathrm{rearrange}}(n)$ in Eq.~\eqref{eq:rearrangement_time} is a function of the code size $n$, as $L$ is a function of $n$. Assuming the idling errors $p_i(n)$ can improve together with the gate error $p_g$ as hardware improves, we set 
\begin{equation}
    p_i(n) = t_{\mathrm{rearrange}}(n)/T_c \times p_g/0.005,
    \label{eq:idling_error_expression}
\end{equation}
where $T_c$ is the atom coherence time and 0.005 is the current CZ gate fidelity demonstrated in Ref.~\cite{evered2023high}. For all the numerical estimations in this work, we use the upper bound for $t_{\mathrm{rearrange}}(n)$ in Eq.~\eqref{eq:rearrangement_time} and use the experimental parameters listed below Eq.~\eqref{eq:rearrangement_time}, with $T_c = 10$s. 

\begin{figure}
    \centering
    \includegraphics[width=0.5\textwidth]{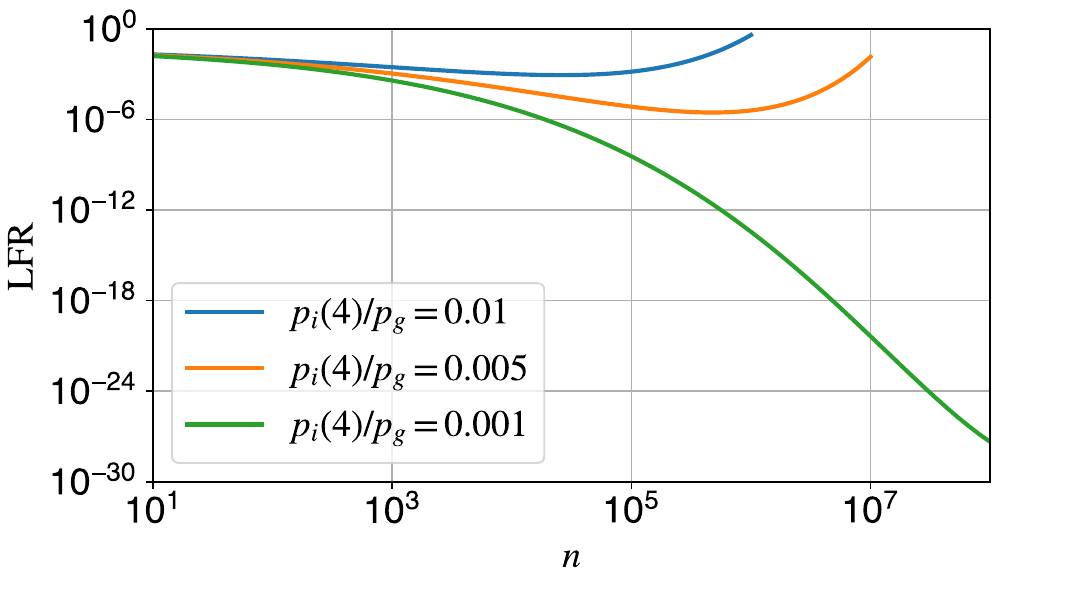}
    \caption{\textbf{Achievable logical failure rates of the HGP codes with different idling error strengths.} We characterize the idling error strengths as the relative ratio between the idling error rate $p_i(n)$ at $n = 4$ and the gate error rate $p_g$. This idling error strength can potentially be reduced by, e.g., increasing the coherence time and accelerating the atom shuttling. }
    \label{fig:LFR_limited_idling}
\end{figure}

In Supplement~\cite{SM}, we numerically verify that the effect of the idling errors can be approximated by rescaling the gate errors $p_g \rightarrow p_g + 3p_i(n)$ using the product coloration circuit. Replacing the rescaled $p_g$ in the subthreshold scaling for HGP codes in Eq.~\eqref{eq:subthreshold_scaling}, we can examine the effect of the idling errors on achievable logical failure rates. As shown in Fig.~\ref{fig:LFR_limited_idling}, the LFRs are first exponentially suppressed by the code size $n$ when $n$ is small and $3p_i(n) \ll p_g$, and gradually saturate and then increase as $3p_i(n) > p_g$ and finally approaches the gate error threshold. Using the relevant experimental parameters, the idling errors are negligible for $n$ up to $\sim 10^7$ and the LFRs can go below $10^{-24}$ (see the green curve), which already suffices for implementing practical quantum algorithms.
Note that for even larger sizes, concatenation with another code can be employed to extend the effective coherence time~\cite{pattison2023hierarchical,raveendran2022finite} and further suppress $p_i(n)$.

Note that our syndrome extraction circuit is not $2D$-local, since coupling two qubits does not incur an error proportional to their Euclidean distance. The effective syndrome extraction circuit depth is $\Delta[1 + 3p_i(n)/p_g]$, where the constant $\Delta$ is the entangling gate depth.  
In the regime where $3p_i(n) \ll p_g$, the circuit is quasi-nonlocal, with a constant circuit depth $\Delta$, whereas in the regime where $3p_i(n) \gg p_g$, the circuit is effectively 4D-local for the HGP codes according to the bound on the circuit depth in finite dimensions in Ref.~\cite{delfosse2021bounds}. Thus, although with the inclusion of idling errors, the logical failure rate cannot be arbitrarily suppressed asymptotically according to the no-go theorems in Ref.~\cite{delfosse2021bounds} and Ref.~\cite{baspin2023lower}, constant overhead and fault tolerance can still be achieved at physically relevant sizes by utilizing the quasi-nonlocal connectivity in atom arrays.

\noindent\textbf{Decoder}\\
Here, we provide details on the circuit-level space-time decoder used in our work. Our decoder is based on the belief-propagation and ordered-statistical decoding (BP+OSD) algorithm proposed in Ref.~\cite{panteleev2019degenerate}. 
See Ref.~\cite{panteleev2019degenerate} and Ref.~\cite{roffe2020decoding} for an introduction to generic BP+OSD decoders. 

When applying consecutive QEC cycles, we use the BP decoder on intervals of noisy cycles to control error accumulation. For the final round, which typically has a noiseless readout, we apply the BP+OSD decoder to eliminate residual syndromes and project the system into the codespace. To decode noisy syndromes using the BP decoder, we construct a space-time circuit-level decoding graph. This allows our BP decoder to decode over multiple QEC cycles while taking circuit-level details into account. Note that a similar circuit-level BP decoder was used in Ref.~\cite{higgott2023improved1} to improve the matching graph for surface code decoding using the minimum-weight perfect matching decoder, and similar circuit-level decoding graphs were used for correcting a generic stabilizer circuit in recent work~\cite{delfosse2023spacetime}. For the final BP+OSD decoder, we use a simple graph with only data errors, as it decodes noiseless syndromes.

Now, we provide details on the construction of the space-time circuit-level decoding graph for the BP decoder. Given a code, we consider a QEC circuit consisting of multiple code cycles. Each code cycle includes the initialization of ancillae, a sequence of parallel entangling gates between ancillae and data qubits, and measurements on the ancillae. Each ancilla measurement in the $t$-th cycle produces a stabilizer measurement outcome, denoted as $S_i^{(t)}$ (for the $i$-th ancilla). We use the notion of detectors in Stim~\cite{gidney2021stim}, defined as the pairty of consecutive stabilizer measurement outcomes: $D_i^{(t)} := S_{i}^{(t)} + S_{i}^{(t - 1)} (\textrm{mod}\;2)$. We then construct a bipartite decoding graph over $T$ code cycles. Check nodes are associated with detectors, and bit nodes are associated with circuit faults (e.g., two-qubit Pauli errors for a CNOT gate) for $t=1,2,\cdots,T$. The connectivity between check nodes and bit nodes is naturally determined by how circuit faults trigger detectors.
The prior error probabilities (used by the BP decoder) for bit nodes are given by the circuit-level noise model used in the simulation. See Ref.~\cite{higgott2023improved} for more details about the decoding graph. Here, to account for residual data errors after the previous single-shot decoding round, we add a layer of depolarizing data errors before the first code cycle to the decoding graph. The error prior for these data errors is currently set phenomenologically as a hyperparameter. See Supplement~\cite{SM} for an example decoding graph for a length-3 repetition code.
For the simulations in this work, we use a min-sum variant of the BP decoder~\cite{emran2014simplified} and an order-$10$ OSD decoder in Ref.~\cite{panteleev2019degenerate}. For the min-sum BP decoder, we set the maximum iteration number to be $n/R$, where $n$ is the size of the code that the decoder decodes on and $R$ is numerically optimized. We also numerically optimize the scaling factor $s$ of the min-sum decoder and the phenomenological error prior $p_r$ (normalized by the gate error rate $p_g$) for the virtual nodes associated with the residual data errors in the decoding graph. We choose $R = 5, s = 0.9, p_r = 5$ for the HGP codes and $R = 20, s = 1, p_r = 1$ for the LP codes. We use Stim~\cite{gidney2021stim} for simulating Clifford circuits and adapt part of the BP+OSD package~\cite{roffe2020decoding, Roffe_LDPC_Python_tools_2022} for decoding.
As described in Ref.~\cite{SM}, our use of space-time circuit level decoding improves the threshold for HGP codes under standard depolarizing noise from below 0.23$\%$ to 0.33$\%$ when using the coloration circuit~\cite{delfosse2021bounds}.

\noindent\textbf{Details of teleportation}\\
Here, we describe our teleportation scheme between the qLDPC code and the surface code. We select a logical $\bar{X}_1$ operator for the qLDPC code of minimum weight. This ensures it contains no sublogicals, i.e. inequivalent logical operators contained in its support, so that the scheme is fault-tolerant under data errors (see Supplement~\cite{SM} for proof).
We then associated the qubit support of $\bar{X}_1$ of the qLDPC code ($\bar{Z}_2$ of the surface code) to the bits of a classical code $C_1$ ($C_2$), and associated the $Z$ ($X$) stabilizers of the qLDPC (surface) code with support on $\bar{X}_1$ ($\bar{Z}_2$) to the checks of $C_1$ ($C_2$). Denoting $H^1$ ($H^2$) as the check matrix for $C_1$ ($C_2$), $H^1_{ij} (H^2_{ij}) = 1$ if the $i$-th $Z$ ($X$) stabilizer checks the $j$-th qubit of $\bar{X}_1$ ($\bar{Z}_2$).
We construct an ancilla code patch as a HGP code taking the hypergraph product of $C_1$ and $C_2$, which encodes a single logical qubit with a logical $X$ and $Z$ representative associated with the bits of $C_1$ and $C_2$, respectively.

Lattice surgery between the ancilla patch and the qLDPC (surface) code is realized by merging and splitting along $C_1$ ($C_2$), assisted by an extra array of ancillary qubits (see Fig.~\ref{fig:teleportation}(a)). For a classical linear code $C$ with check matrix $H$, denote by $C^T$ its transposed code defined by the check matrix \(H^T\).
Taking the qLDPC-ancilla surgery as an example,  we insert an extra array of $X$ stabilizers (initialized in $\ket{+}$) and qubits (initialized in $\ket{0}$) in the middle, associated with the checks and the bits of $C_1^T$, respectively. During the code merging, the Z stabilizers of the qLDPC and the ancilla patch associated with the $i$-th check of $C_1$ are each modified to include the middle qubit associated with the $i$-th bit of $C_1^T$; The middle $X$ stabilizer associated with the $j$-th check of $C_1^T$ checks the two qubits of the qLDPC and surface codes associated with the $j$-th bit of $C_1$ as well as some middle qubits given by the incident relation of $C_1^T$. It is easy to verify that all the new stabilizers commute as the added/modified qubits and checks across the merged boundary form an HGP code with $C_1$ and a length-$2$ repetition code locally.  The product of the middle $X$ stabilizers give the joint logical operator to measure. See Supplement~\cite{SM} for more algebraic details of the above lattice-surgery scheme, and a proof of its fault tolerance under data errors. 

We perform the numerical simulations using Stim~\cite{gidney2021stim}, and use the same space-time decoder described in the previous section for decoding.
For ease of simulation, we add circuit depolarizing noise with no idling errors on the joint logical $XX$ measurement part of the teleportation circuit, in both the merging and splitting steps, and rescale the gate error rate $p_g \rightarrow p_g + 3p_i$ to obtain Fig.~\ref{fig:teleportation}(c).
As the merged code no longer supports single-shot error correction, we perform $d$ rounds of QEC after each of the merge and split steps of the $XX$ surgery, where $d=\min (d_1, d_2)$ is the distance of the teleportation scheme and $d_1$ and $d_2$ are the distances of the qLDPC code and surface code, respectively.
We expect the logical failure rate of the $ZZ$ logical measurement to be nearly the same as that of the qLDPC memory if we use a computation surface code with a distance larger than that of the qLDPC code.
For the logical $M_X$ measurement, we simulate a noiseless destructive measurement.

We found a gate error threshold (without idling errors) of $\sim 0.7\%$ by looking at the crossing of $d=3,5,7$ schemes, which were constructed with $(d_1,d_2) = (4,3),(6,5),(8,7)$, respectively.
We attribute the slight increase in the threshold compared to the memory to the increase in the number of code cycles used by the space-time decoder (the memory simulations are decoded using only three cycles).

Note that the lattice surgery approach in Ref.~\cite{cohen2022low} can also be used for teleportation between a qLDPC code and a surface code. Their approach essentially uses an ancilla patch formed by the hypergraph product of a length-$d$ repetition code and a ``union" of $C_1$ and $C_2$ associated with the logical operators of the qLDPC code and the surface code respectively, and directly performs a joint logical measurement on the qLDPC and surface code. The ancilla patch has size $2d^2$ (if using minimum-weight logical operators), which is twice as larger as our ancilla. Compared to their approach, our scheme has a lower space overhead but a larger temporal overhead overall.

\end{document}


\title{Supplementary Information for ``Constant-Overhead Fault-Tolerant Quantum Computation with Reconfigurable Atom Arrays"}

\author{Qian Xu}
\thanks{These authors contributed equally.}
\affiliation{Pritzker School of Molecular Engineering, The University of Chicago, Chicago 60637, USA}

\author{J. Pablo Bonilla Ataides}
\thanks{These authors contributed equally.}
\affiliation{Department of Physics, Harvard University, Cambridge, Massachusetts 02138, USA}

\author{Christopher A. Pattison}
\affiliation{Institute for Quantum Information and Matter, California Institute of Technology, Pasadena, CA 91125}

\author{Nithin Raveendran}
\affiliation{Department of Electrical and Computer Engineering, University of Arizona, Tucson, AZ 85721, USA}


\author{Dolev Bluvstein}
\affiliation{Department of Physics, Harvard University, Cambridge, Massachusetts 02138, USA}

\author{Jonathan Wurtz}
\affiliation{QuEra Computing Inc., 1284 Soldiers Field Road, Boston, MA, 02135, US}

\author{Bane Vasić}
\affiliation{Department of Electrical and Computer Engineering, University of Arizona, Tucson, AZ 85721, USA}

\author{Mikhail D. Lukin}
\affiliation{Department of Physics, Harvard University, Cambridge, Massachusetts 02138, USA}

\author{Liang Jiang}
\email{liang.jiang@uchicago.edu}
\affiliation{Pritzker School of Molecular Engineering, The University of Chicago, Chicago 60637, USA}

\author{Hengyun Zhou}
\email{hyzhou@quera.com}
\affiliation{Department of Physics, Harvard University, Cambridge, Massachusetts 02138, USA}
\affiliation{QuEra Computing Inc., 1284 Soldiers Field Road, Boston, MA, 02135, US}

\maketitle

\section{Circuit-level fault tolerance}
In this section, we analyze the fault-tolerance of qLDPC codes under circuit-level noise, present the details of the space-time decoder we use, report the numerical circuit-level thresholds of the HGP and the LP codes, and analyze the effect of the idling errors associated with our atom rearrangement scheme. 

\subsection{Threshold theorem}
Here, we gather known results in the literature to show that various qLDPC codes have a single-shot threshold under circuit-level depolarizing noise.
To present the results coherently, we have slightly modified definitions and theorem statements.
We use \([n]\) to denote the set of integers \(\{1, \dots, n\}\), and \(\mathcal{P}\) to denote the Pauli group \(\{X,Y,Z,I\}\).
Proofs of the quoted results may be found in the cited references.

The noise model that we consider is the standard circuit-level depolarizing noise model.
\begin{definition}[Circuit-level depolarizing noise]
  \label{defn:circuit-noise}
  Let \(\mathcal{C}\) be an ideal Clifford circuit containing one and two-qubit gates.
  The noisy circuit \(\tilde{\mathcal{C}}\) under circuit-level depolarizing noise at rate \(p\) contains the gates of the ideal circuit as well as Pauli operators (errors) applied after each gate independently.
  Precisely:
  \begin{enumerate}
  \item After each one-qubit gate (including identity), apply \(I\) with probability \(1-p\) and otherwise one of \(X\), \(Y\), or \(Z\) picked uniformly at random.
  \item After each two-qubit gate, apply \(I\) with probability \(1-p\) and otherwise one of the 15 non-trivial two-qubit Pauli operators picked uniformly at random.
  \item Flip each measurement outcome with probability \(p\).
  \end{enumerate}
\end{definition}

When applied to the single-ancilla syndrome extraction circuit of qLDPC codes, such a noise model falls into a broad class of Pauli noise models known as \emph{local stochastic noise} models.
\begin{definition}[Local stochastic syndrome-data noise]
  \label{defn:local-stochastic-noise}
  Let \(\mathcal{C}\) be a syndrome extraction circuit of a stabilizer code that acts on an \(n\) qubit register and produces \(m\) measurement outcomes.
  We restrict our attention to codestates with added Pauli noise, and model noise as a random variable \((e,x)\) taking values in \(\mathcal{P}^n\times \mathbb{F}_2^m\).
  For an \(n\) qubit input state \(\ket{\psi}\), the corresponding output state is \(e \ket{\psi}\).
  Likewise, letting the noiseless measurement outcomes be \(y \in \mathbb{F}_2^m\), the noisy measurement outcome is \(x+y\).
  We refer to such noise models as \emph{syndrome-data noise}.

  A syndrome data noise model is said to be \emph{\(p\)-local stochastic} if, for any set \(F \subseteq [n+m]\), we have that
  \begin{align}
    \mathbb{P}[F\subseteq \supp(e) \cup \supp(x)] \le p^{|F|}
  \end{align}
\end{definition}

\begin{theorem}[Theorem 5.4 of \cite{pattison2023hierarchical}]
  \label{thm:circuit-is-local-stochastic}
  Fix a CSS qLDPC code with the depth \(\Delta\) syndrome extraction circuit \(\mathcal{C}\) defined in the main text that acts on \(n\) data qubits and provides \(m\) measurement results.
  Circuit-level depolarizing noise at a rate of \(p\) is local stochastic syndrome-data noise at a rate of \(\Delta 2^{\Delta+1} p^{1/(2\Delta + 2)}\).
\end{theorem}
We note that the low-density property of the qLDPC codes is crucial for Theorem~\ref{thm:circuit-is-local-stochastic} to hold since it guarantees that the single-ancilla syndrome extraction circuit used in the main text has constant depth and only permits bounded error propagation.

A sufficient condition for proving the existence of single-shot decoders is known as \emph{confinement}~\cite{quintavalle2020single}.
\begin{definition}[Reduced Weight]
  \label{defn:reduced-weight}
  For a stabilizer code on \(n\) qubits with stabilizer group \(\mathbb{S} \subseteq \mathcal{P}^n\), the reduced weight \(|\cdot|_{R}\) of a Pauli operator \(x \in \mathcal{P}^n\) is defined to be
  \begin{align}
    |x|_R :=\min_{s \in \mathbb{S}} |xs|,
  \end{align}
  where \(| \cdot |\) denotes the size of the support of the Pauli operator.
\end{definition}
\begin{definition}[Confinement]
  \label{defn:confinement}
  A quantum code on \(n\) qubits with syndrome map \(\sigma \colon \mathcal{P}^n \to \mathbb{F}_2^r\) has \((\gamma, \alpha)\)-linear confinement, if, for any Pauli operator \(E \in \mathcal{P}^n\) such that \(|E|_R \le \gamma\), the size of the syndrome is proportional to the size of the error: \(|\sigma(E)|_R \ge \alpha |E|_R\). A code family has \((\gamma, \alpha)\)-linear confinement if each code of size $n_i$ has a $(\gamma(n_i), \alpha)$-linear confinement where $\gamma(n_i) = \Omega(\mathrm{poly}(n_i))$.
\end{definition}

It was shown in \cite{quintavalle2020single} that for any qLDPC code family satisfying the confinement property, a (not necessarily efficient) single-shot decoder with a threshold against local stochastic syndrome-data noise exists.

\begin{theorem}[Theorem 2 of \cite{quintavalle2020single}]
  \label{thm:single-shot-thresh}
  For a family of qLDPC codes indexed by \(i\) on \(n_i\) qubits with \((\gamma_i, \alpha)\)-linear confinement such that \(\gamma_i = \Omega(\mathrm{poly}(n_i))\), under a model of \(p\)-local stochastic syndrome-data noise, there exists a single-shot decoder for which the code family has a sustainable threshold. I.e. there exists a \(p^* \in (0,1)\) such that for \(p < p^*\), after any constant number of faulty syndrome extraction and correction cycles, the probability the data block is unsuccessfully recovered by an ideal recovery map is exponentially small in \(\mathrm{poly}(n_i)\).
\end{theorem}

Combining Theorem~\ref{thm:circuit-is-local-stochastic} and Theorem~\ref{thm:single-shot-thresh} suffices to prove that a single-shot circuit-level threshold exists for qLDPC codes satisfying confinement using the single-ancilla syndrome extraction circuit in the main text. However, it is not easy to check whether or not a generic quantum code satisfies the confinement property. Fortunately, the existence of small-set-flip-type decoders for a code family provides the required property as a short corollary.
We begin with a formal definition of such a class of decoders:
\begin{definition}[Local Greedy Decoder]
  Fix a stabilizer code \(\mathcal{C}\) on \(n\) qubits with syndrome map \(\sigma \colon \mathcal{P}^n \to \mathbb{F}^r_2\) and an error \(E \in \mathcal{P}^n\).
  Consider a decoder that proceeds iteratively in \(t\) steps, where for each step \(i \in [t]\), the syndrome weight is strictly decreased by adding a correction \(C_i \in \mathcal{P}^n\) to the current algorithm state \(E + C_1 + \dots C_{i-1}\).
  If the decoder successfully corrects all errors of reduced weight at most \(s\) and the size of each correction is at most \(c\), then the decoder is said to be a \((s, c)\)-local greedy decoder.
\label{def:local_greedy_decoder}
\end{definition}

Using this definition, we can prove that codes with such decoders possess linear confinement.
\begin{lemma}[Local greedy decoders imply confinement]
  \label{lemma:greedy-decoders}
  If there exists an \((s,c)\)-local greedy decoder for a code, then it possesses \((s, 1/c)\)-linear confinement.
\end{lemma}
\begin{proof}
  The decoder applies a correction of size at most \(c\) in each of \(t\) updates, so the total correction \(C\) is of size at most \(ct\).
  By assumption, if \(|E|_{R}\le s\) the decoder corrects successfully \(|E+C|_R = 0\), so \(|E|_R \le |C| \le ct\) (otherwise, we arrive at a contradiction with the definition of reduced weight).
  
  On the other hand, at each step the syndrome weight is decreased by at least one, so \(|\sigma(E)| \ge t\).
  We conclude that \(|\sigma(E)|\ge \frac{1}{c} |E|_R\) for all errors satisfying \(|E|_R\le s\).
\end{proof}

We are now ready to conclude with the final threshold theorem:
\begin{theorem}[qLDPC codes permitting local greedy decoders are single-shot decodable]
    Consider a family of quantum codes indexed by \(i \in \mathbb{N}\) with each element of the family encoding into \(n_i\) qubits with \(n_i\) increasing with \(i\).
    If each element of the family permits a \((s_i,c)\)-local greedy decoder with \(s_i = \Omega(\mathrm{poly}(n_i))\), then there exists a single-shot decoder for which the code family has a sustainable threshold under circuit level depolarizing noise using the syndrome extraction circuit from the main text.
\end{theorem}
\begin{proof}
    By assumption, the \(i\)-th element of the code family permits an \((s_i, c)\)-local greedy decoder, so \cref{lemma:greedy-decoders} implies each element also has \((s_i, 1/c)\)-linear confinement.
    \cref{thm:circuit-is-local-stochastic} shows that circuit-level depolarizing noise applied to a constant-depth syndrome extraction circuit is local-stochastic syndrome-data noise, so the preconditions of \cref{thm:single-shot-thresh} are met.
\end{proof}

Many code families based on expander graphs are known to have local greedy decoders as expansion properties frequently allow a greedy local search.
This includes, for appropriate instantiations of the parameters, Quantum Expander codes \cite{ leverrier2015quantum,tillich2014quantum}, Pantaleev-Kalachev codes \cite{panteleev2022asymptotically,leverrier2022quantum}, Quantum Tanner Codes \cite{leverrier2022quantum,gu2022efficient,leverrier2022efficient}, the construction of Dinur-Hseih-Lin-Vidick on square complexes \cite{dinur2022good}, codes based on symmetric lossless expanders \cite{lin2022good} and codes based on simplicial Ramanujan Complexes \cite{evra2020decodable}.
We remark that better parameters can frequently be obtained by direct analysis, but tight bounds are often very difficult to prove, limiting the value of parameter optimization. Note that the existence proof of the single-shot thresholds here utilizes a stochastic shadow decoder~\cite{quintavalle2020single}, which is not efficient in practice.
With appropriate choices of parameters, efficient single-shot decoders are known for some constant rate code constructions such as hypergraph product codes \cite{fawzi2018constant}, Quantum Tanner codes \cite{gu2023single}, and Pantaleev-Kalachev codes \cite{panteleev2022asymptotically}.

\begin{figure*}
    \centering
    \includegraphics[width=1\textwidth]{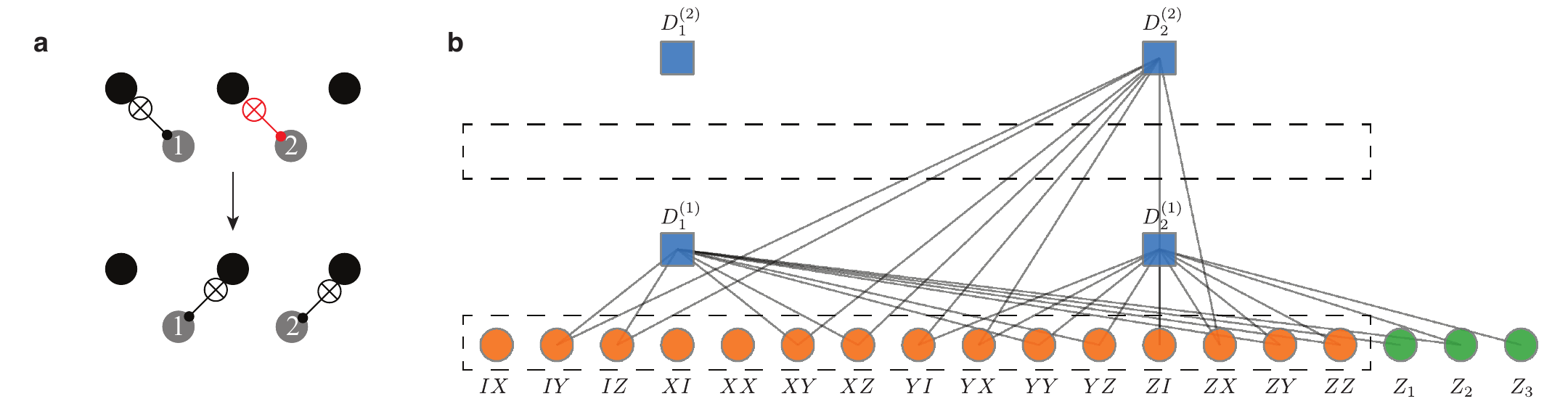}
    \caption{\textbf{Example of the space-time decoding graph for a length-3 repetition code correcting $Z$ errors.} (a) The syndrome extraction circuit during a code cycle. (b) A subset of the decoding graph consisting of the faults associated with the highlighted CNOT gate in (a) in the first code cycle (the orange circles) and the phenomenological data faults (the green circles). }
    \label{fig:decoding_graph}
\end{figure*}

Here, we provide an example of the circuit-level space-time decoding graph that we use for the BP decoder. We consider a length-$3$ repetition code, whose syndrome extraction circuit during one code cycle is illustrated in Fig.~\ref{fig:decoding_graph}(a). In Fig.~\ref{fig:decoding_graph}(b) we plot a subset of its decoding graph, which only includes the faults associated with the CNOT gates between the second check ancilla and the middle qubit in the first code cycle (see the highlighted CNOT gate in Fig.~\ref{fig:decoding_graph}(a) and the orange faults nodes in Fig.~\ref{fig:decoding_graph}(b)). In addition, we also show the phenomenologically added data faults before the first code cycle (see the green circles). These faults will trigger the detectors associated with the two check ancillae (represented by the blue squares) in the first and second code cycles.

\subsection{Numerical simulations}
We conduct memory simulations by initializing logical qubits,  simulating $m \gg 1$ syndrome extraction cycles, and performing a transversal qubit readout at the end. We use Stim~\cite{gidney2021stim} to sample the above circuit. A logical failure is recorded if any of the logical qubits have a flipped logical observable after the decoding. Denoting the above total logical failure probability as $p_L$, we obtain the logical failure rate per code cycle as $\mathrm{LFR} = 1 - (1 - p_L)^{1/m}$. We calculate the standard deviation of $p_L$ using the standard formula for binomial distributions $\sigma_{p_L} = \sqrt{p_L(1 - p_L)/N}$, where $N$ denotes the number of samples used in the Monte Carlo simulation. Then we obtain the standard deviation of $\mathrm{LFR}$, which are represented by the error bars in our $\mathrm{LFR}$ plots, via the error propagation formula $\sigma_{\mathrm{LFR}} = |\partial \mathrm{LFR}/\partial p_L|\sigma_{p_L} = (1 - p_L)^{1/m - 1}\sigma_{p_L}$. We choose $N$ adaptively, ranging from a few hundred to $\sim 10^5$, for different codes and physical error rates to minimize computational resources while keeping $\sigma_{\mathrm{LFR}}$ small.

\begin{figure}
    \centering
    \includegraphics[width=1\textwidth]{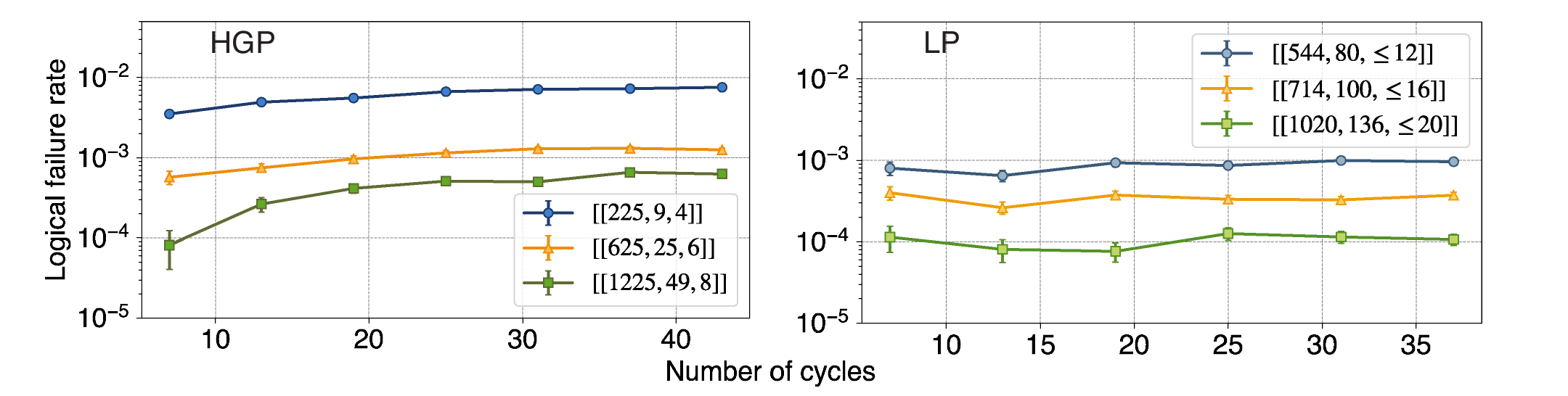}
    \caption{\textbf{Verification of the single-shot property of the HGP codes and the LP codes.} The logical failure rate (per code cycle) stabilizes as the number of code cycles increases for both the HGP codes and the LP codes. The physical error rate is fixed at $0.15\%$ for the simulations. }
    \label{fig:single_shot}
\end{figure}

We first verify that the HGP codes and the LP codes can indeed be single-shot decoded using our space-time BP+OSD decoder numerically. 
We apply the space-time BP decoder over every three code cycles (regardless of the code sizes) and the BP+OSD decoder to the final cycle.
In Fig.~\ref{fig:single_shot}, we show that the logical failure rate converges to a stable value as we increase the number of code cycles, thus verifying the single-shot property of the HGP and the LP codes using our space-time decoder.

\begin{figure}
    \centering
    \includegraphics[width=0.5\textwidth]{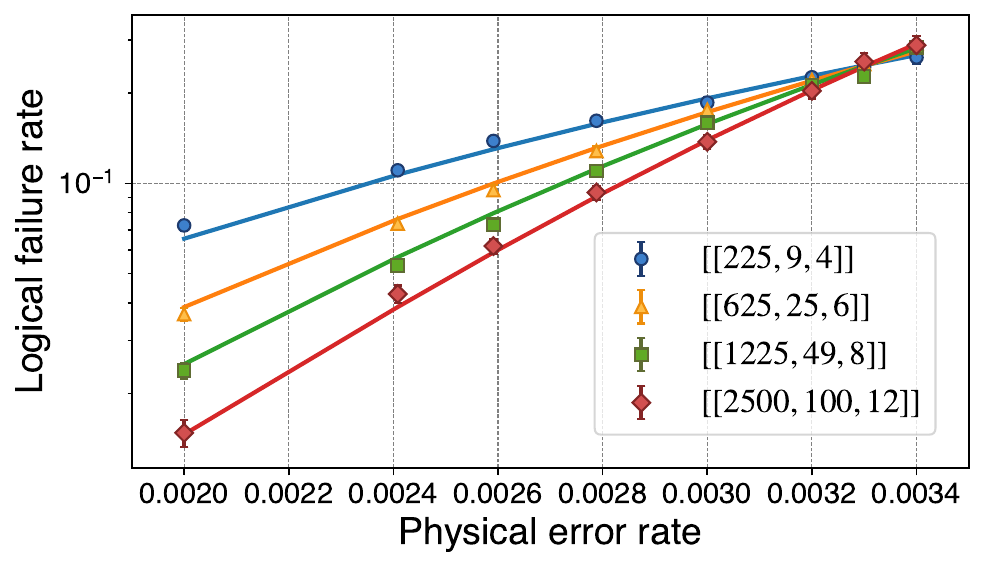}
    \caption{\textbf{Logical failure rates of the HGP codes using the space-time decoder under the standard depolarizing noise model.} The threshold is $0.33\%$, which is higher than that ($< 0.23\%$) using a phenomenological decoder~\cite{delfosse2021bounds}. The number of code cycles for the simulation is $12$. }
    \label{fig:depolarizing_threshold}
\end{figure}

Now, we report the numerical thresholds of the HGP codes and the LP codes. First, we show that our space-time decoder can improve upon the phenomenological decoder used in the literature~\cite{tremblay2022constant, delfosse2021bounds}. We consider the same HGP code family with the same coloration syndrome extraction circuit under the same standard depolarizing noise model as in Ref.~\cite{delfosse2021bounds}. For all threshold estimation in this work, we fit near-threshold data to a variation of the critical-exponent expression~\cite{wang2003confinement}
\begin{equation}
    \mathrm{LFR} = A + B x + C x^2,
\end{equation}
where $x := (p - p_c) \alpha n^{\beta}$, $p_c$ is the threshold to be estimated, and $n$ is the code size. As shown in Fig.~\ref{fig:depolarizing_threshold}, we find a threshold of $0.33\%$, which is higher than that ($< 0.23\%$) reported in Ref.~\cite{delfosse2021bounds}. We note that the subthreshold scaling is also greatly improved by using our space-time circuit-level decoder. 

\begin{figure}[h!]
    \centering
    \includegraphics[width=1\textwidth]{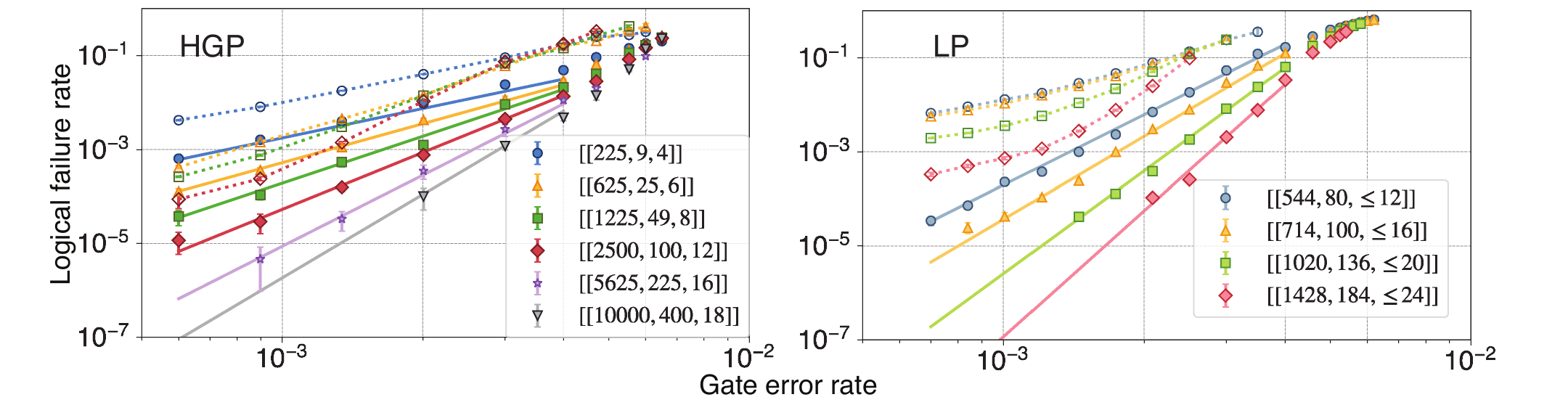}
    \caption{\textbf{Logical failures of the HGP codes and the LP codes when only adding two-qubit gate errors.} The solid lines represent the fitted subthreshold scaling: $\textrm{LFR} = 0.07(p_g/0.006)^{0.47n^{0.27}}$ for the HGP codes and $\textrm{LFR} = 2.3(p_g/0.0066)^{0.11n^{0.6}}$ for the LP codes. For the HGP codes, we simulate $42$ code cycles for $p_g \leq 4\times 10^{-3}$ and $12$ cycles for $p_g > 4\times 10^{-3}$; For the LP codes, we simulate $60$ cycles for $p_g \leq 4\times 10^{-3}$ and $12$ cycles for $p_g > 4\times10^{-3}$. The empty markers along with the dashed lines represent the results using a simpler phenomenological decoder that uses a decoding graph with phenomenological data and measurement errors~\cite{delfosse2021bounds, tremblay2022constant}.
    }
    \label{fig:memory_thresholds}
\end{figure}

Next, we evaluate the logical failure rates of the HGP codes and the LP codes when only considering the depolarizing noise model with no idling errors. The results are shown by the solid markers in Fig.~\ref{fig:memory_thresholds}. The fitted threshold for the HGP codes and the LP codes are $(0.63 \pm 0.01)\%$ and $(0.62 \pm 0.02)\%$, respectively. 
For gate error rates below $4\times10^{-3}$, we numerically fit the subthreshold scaling using the logarithm of the ansatz 
\begin{equation}
\mathrm{LFR} = A (p_g/p_0)^{\alpha n^{\beta}/2}.
\label{eq:subthreshold_ansatz}
\end{equation}
For the HGP codes, we obtain 
$A = 0.07\pm0.04$, $p_0 = 0.60\%\pm0.08\%$, $\alpha = 0.94 \pm 0.21$, and $\beta = 0.27\pm 0.03$; For the LP codes, we obtain $A = 2.3\pm 1.2$, $p_0 = 0.66\%\pm0.06\%$, $\alpha = 0.22 \pm 0.07$, and $\beta = 0.60\pm 0.05$.
We also plot the logical failure rates using the phenomenological decoder (the dashed lines with empty markers) as a comparison. Clearly, the space-time decoder improves not only the thresholds but also the subthreshold scaling. 

\subsection{Effect of idling errors}
\begin{figure}
    \centering
    \includegraphics[width=0.5\textwidth]{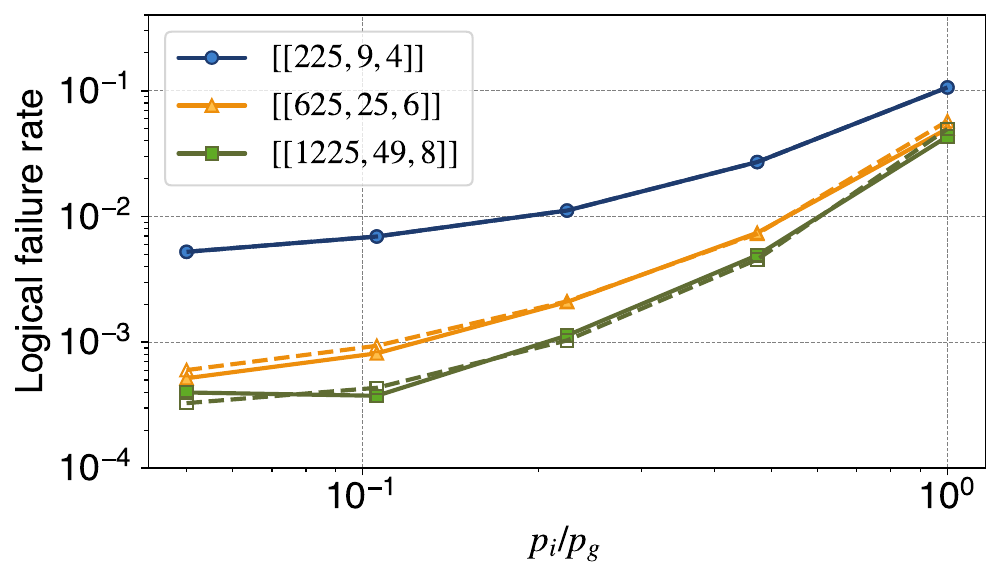}
    \caption{\textbf{Logical failure rates versus the ratio between the idling error $p_i$ and the gate error $p_g$ for the HGP codes.} The gate error rate $p_g$ is set to be $10^{-3}$. The solid lines represent the results when both the gate errors and the idling errors are added to the simulation, while the dashed lines represent the results when only gate errors are added with an error rate $p_g + 3p_i$. }
    \label{fig:Idling_Error}
\end{figure}

Here, we investigate the effect of idling errors between the sequence of parallel CNOT gates connecting check qubits and data qubits. In Fig.~\ref{fig:Idling_Error}, we compare the logical failure rates when both gate errors (with a rate $p_g$) and idling errors (with a rate $p_i$) are added to the simulation (solid lines) and when only gate errors (with a rate $p_g + 3p_i$) are added (dashed lines). The logical failure rates in both cases are similar, confirming that the effect of idling errors can be well approximated by rescaling the gate error rate to $p_g \rightarrow p_g + 3p_i$.

\section{Teleportation}
\subsection{Teleportation of a single logical qubit}
\begin{figure}
    \centering
    \includegraphics[width=0.5\textwidth]{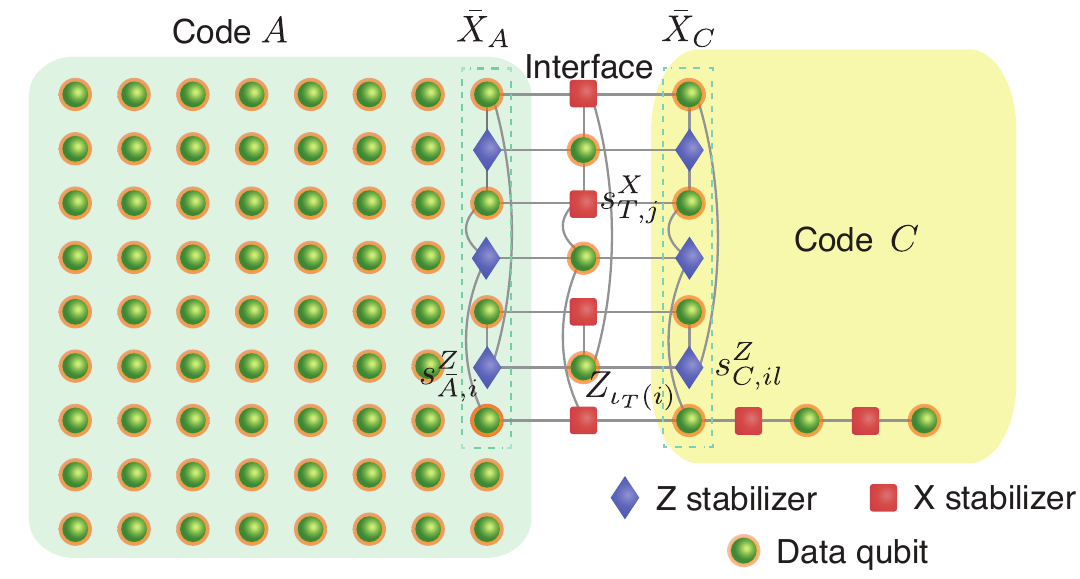}
    \caption{\textbf{Illustration of a lattice surgery between two quantum codes A and C for measuring the joint logical operator $\bar{X}_A \bar{X}_C$. }}
    \label{fig:lattice_surgery_proof}
\end{figure}
We begin by defining the necessary ingredients for the teleportation.
Consider two codes, defined by the \(CSS\) stabilizer generators \(S_A\) and \(S_B\) supported on qubit sets \(A\) and \(B\) respectively.
Let \(\overline{X}_A\) be a minimum weight \(X\) logical operator of the \(A\) code and \(\overline{Z}_B\) be a minimum weight \(Z\) logical operator of the \(B\) code.
We will prove fault tolerance of this code against adversarial data noise by considering a related subsystem code \cite{poulsen2017fault,vuillot2019code} and proving that the dressed distance is large.%

In the following, we will mostly consider the \(A\) patch, but all results generalize to the \(B\) patch by appropriate exchanges of \(X\)/\(Z\) and \(A\)/\(B\).

We will use a superscript to refer to the subset of \(X\) or \(Z\) stabilizer generators, i.e. \(S_A^X = \{s \in S_A \mid s\text{ is a product of Pauli }X\}\) and \(S_A = S^X_A \cup S^Z_A\).
For a qubit set \(W\cong [n]\), we will use \(\iota_A \colon [n] \to W\) to denote the coordinate function into that set.
Occasionally, we will omit the use of the coordinate function when it is clear from context.
For a subset \(T \subseteq W\) and a Pauli operator \(x\in \mathcal{P}^W\) supported on \(W\), we use \(x |_T\) to denote the Pauli operator that is equal to \(x\) on all qubits inside of \(T\) and identity on all qubits outside of \(T\).

\begin{definition}[HGP Code]
  Let \(H_{A}\colon \F^{n_A}\to \F^{r_A}\), and \(H_{B}\colon \F^{n_B} \to \F^{r_B}\) be check matrices over \(F_{2}\).
  The hypergraph product code associated to \((H_A, H_B)\) has two sets of qubits \(CC \simeq [r_A] \times [r_B]\) and \(VV \simeq [n_A] \times [n_B]\).
  The check matrices of the quantum code are \(H_Z = \begin{pmatrix} I_{r_A}\otimes H_B^T & H_A \otimes I_{n_B}\end{pmatrix}\) and \(H_X = \begin{pmatrix}H_A^T \otimes I_{r_B} & I_{n_A} \otimes H_B\end{pmatrix}\).
\end{definition}

\begin{definition}[Ancilla Patch]
\label{def:ancilla_patch}
  The ancilla patch for the tuple \((S_A, S_B, \overline{X}_A, \overline{Z}_B)\) is constructed as follows:
  Let \(m_A = |\supp \overline{X}_A|\) and \(m_B = |\supp \overline{Z}_B|\), and define the restricted stabilizers by the restriction of the stabilizer generators to the support of the selected logical operator.
  \begin{align}
    \bar{S}_A &:= \left\{s|_{\supp \overline{X}_A}\mid (s \in S_A) \;\mathrm{and}\; (\supp s \cap \supp \overline{X}_A \ne \emptyset)\right\} \\
    \bar{S}_B &:= \left\{s|_{\supp \overline{Z}_B}\mid (s \in S_B) \;\mathrm{and}\; (\supp s \cap \supp \overline{Z}_B \ne \emptyset)\right\} 
  \end{align}
  The restricted stabilizers induce classical check matrices given by placing the \(r_{\bar{A}}\) (\(r_{\bar{B}}\)) elements of \(\bar{S}_A^Z\) (\(\bar{S}_B^X\)) as rows of a matrix \(H_{\bar{A}}\) (\(H_{\bar{B}}\)) acting on \(\F^{m_A}\) (\(\F^{m_B}\)).
  The ancilla patch is then the hypergraph product of these classical codes with stabilizers supported on the qubits \(VV \cong [m_A]\times [m_B]\) and \(CC \cong [r_{\bar{A}}]\times [r_{\bar{B}}]\).
  Using \(A[i,j]\) to denote the entry of a matrix \(A\) in the \(i\)-th row and \(j\)-th column, the stabilizer generators of the ancilla patch are \(S_C^Z = \{s^Z_{i,j}\}_{(i,j) \in [r_{\bar{A}}] \times [m_B]}\), \(S_C^X = \{s^X_{i,j}\}_{(i,j) \in [m_A] \times [r_{\bar{B}}]}\) where
  \begin{align}
    (i,j) \in [r_{\bar{A}}] \times [m_B]&\; &s_{i,j}^Z = \left(\prod_{(k,\ell) \in VV, \ell = j} Z_{k,\ell}^{H_{\bar{A}}[i,k]}\right)\left(\prod_{(k,\ell) \in CC, k = i} Z_{k,\ell}^{H_{\bar{B}}^T[j,\ell]}\right)\\
    (i,j) \in [m_A] \times [r_{\bar{B}}]&\; &s_{i,j}^X = \left(\prod_{(k,\ell) \in VV, k = i} X_{k,\ell}^{H_{\bar{B}}[j,\ell]}\right)\left(\prod_{(k,\ell) \in CC, \ell=j} X_{k,\ell}^{H_{\bar{A}}^T[i,k]}\right)
  \end{align}
  We will use the coordinates of these stabilizer generators later.

  We refer to an ancilla patch as a non-degenerate ancilla patch when \(H_{\bar{B}}^T\) has trivial kernel and both \(\overline{X}_A\), \(\overline{Z}_B\) are minimal weight logical operators.
\end{definition}

\begin{remark}
  A minimal weight logical operator \(\overline{Z}_{B}\) for which \(H_{\bar{B}}\) has trivial kernel can be easily found when the code is a surface code or color code with a boundary.
  Then, \(H_{\bar{B}}\) is the full rank check matrix for the repetition code with row and column weight at most 2.
\end{remark}

\begin{proposition}
    \label{prop:ancilla-patch-params}
    The non-degenerate ancilla patch for \((S_A, S_B, \overline{X}_A, \overline{Z}_B)\) has one logical qubit and distance \(\min\left(|\overline{X}_A|, |\overline{Z}_B|\right)\).
    Furthermore, minimal weight representatives of \(\overline{Z}\) (\(\overline{X}\)) are given by \(Z\) (\(X\)) supported on rows (columns) of \(VV\).
\end{proposition}

\begin{proof}
  The ancilla patch is a hypergraph product code, and \(\ker H_{\bar{B}}^T\) is trivial by assumption.
  
  First note that any element of \(\ker H_{\bar{A}}\) must also be an element of \(\ker H_A\), so the existence of a non-trivial element of \(\ker H_{\bar{A}}\) with weight strictly less than \(|\overline{X}_A|\) would contradict the minimality of \(\overline{X}_A\).
  Additionally, by construction, elements of \(\ker H_{\bar{A}}\) are bit strings of length at most \(\ker H_{\bar{A}}\), so we also conclude that \(\ker H_{\bar{A}}\) is a set of size two containing the all-ones vector and the zero vector.
  An identical argument holds for \(H_{\bar{B}}\).

  We conclude that \(\ker H_{\bar{A}}\) and \(\ker H_{\bar{B}}\) encode \(k_A=k_B=1\) bit.
  By assumption \(\ker H_{\bar{B}}^T\) encodes \(k_B^T = 0\) bits.
  Let \(\log_2 |\ker H_{\bar{A}}^T| = k_A^T\).
  Using the standard rate formula \cite{tillich2014quantum}, the hypergraph product code encodes \(k_A k_B + k_A^T k_B^T = 1\) qubits.

  We now proceed to the distance:
  Since the all-1s vector is in the kernel of \(H_{\bar{A}}\) and \(H_{\bar{B}}\), it then follows that each element in the sets \(\left\{\prod_{k\in{[m_A]}} X_{\iota_{VV}(k,\ell)}\right\}_{\ell \in [m_B]}\) and \(\left\{\prod_{\ell\in{[m_B]}} Z_{\iota_{VV}(k,\ell)}\right\}_{k \in [m_A]}\) are representatives for the logical operators \(\overline{X}\) and \(\overline{Z}\), respectively.
  Since a representative of \(\overline{X}\) (\(\overline{Z}\)) must anti-commute with every representative of \(\overline{Z}\) (\(\overline{X}\)), it must share support with every representative of \(\overline{Z}\) (\(\overline{X}\)).
  Since we have \(\min(m_A, m_B)\) representatives with disjoint supports for every non-trivial logical operator, we conclude that the minimum weight of any representative must be at least \(\min(m_A, m_B) \equiv \min\left(|\overline{X}_A|, |\overline{Z}_B|\right)\).
\end{proof}

\begin{definition}[Merged Patch]
\label{def:merged_patch}
  Fix an arbitrary coordinate \(\ell \in [m_B]\) \footnote{This fixes a choice of logical representative of the ancilla patch.}.
  We define the \((A,\ell)\)-merged patch between the code defined by \(S_A\) and the \((S_A, S_B, \overline{X}_A, \overline{Z}_B)\) ancilla patch in the following way: Introduce a set of new qubits with one for each row of \(H_{\bar{A}}\), \(T \cong [r_{\bar{A}}]\).
  We refer to the set \(T\) as interface qubits.
  For \(i \in [r_{\bar{A}}]\), let \(s_{\bar{A},i}^Z\in S^Z_{\bar{A}}\) refer to the element of \(S^Z_{\bar{A}}\) associated with the \(i\)-th row of \(H_{\bar{A}}\).
  Likewise, for \((i,j) \in [r_{\bar{A}}] \times [m_B]\), let \(s_{C,ij} \in S^Z_C\) refer to the stabilizer generators defined with the earlier coordinates.
  The merged patch has the stabilizer generators \(S_M\) given by the union of:
  \begin{align}
    i \in [m_A]&\;& S^X_{T,i} := X_{\iota_{\bar{A}}(i)}X_{\iota_C(i, \ell)} \prod_{j\in [r_{\bar{A}}]} X_{\iota_T(j)}^{H_{\bar{A}}^T[i,j]}\label{eq:A-merged-stabilizers-T}\\
    i \in [r_{\bar{A}}] &\;& s_{\bar{A},i}^ZZ_{\iota_T (i)} \label{eq:A-merged-stabilizers-2}\\
    i \in [r_{\bar{A}}] &\;& s^Z_{C,i\ell} Z_{\iota_{T}(i)} \label{eq:A-merged-stabilizers-3}\\
    (i,j) \in [r_{\bar{A}}] \times [m_B],\; j\ne \ell &\;& s^Z_{C,ij} \label{eq:A-merged-stabilizers-4}\\
    &S_A\setminus S_{\bar{A}}^Z& \label{eq:A-merged-stabilizers-5}\\
    &S_C^X \label{eq:A-merged-stabilizers-6}&
  \end{align}
\end{definition}
See Fig.~\ref{fig:lattice_surgery_proof} for an illustration of the merged patch and the merged stabilizers.

\begin{proposition}\label{prop:merged-params}
  Let \(\overline{X}_C\) denote the \(X\) logical of the \((S_A, S_B, \overline{X}_A, \overline{Z}_B)\) ancilla patch.
  The stabilizer generators of the \((A,\ell)\)-merged patch mutually commute, and furthermore, the joint logical operator is in the stabilizer group of the \((A,\ell)\)-merged patch i.e. \(\langle S_M\rangle\) is an abelian group and \(\overline{X}_C \overline{X}_A \in \langle S_M\rangle\).
\end{proposition}
\begin{proof}
The only non-trivial pairs to check are the stabilizers generators from \cref{eq:A-merged-stabilizers-T} with those of \cref{eq:A-merged-stabilizers-2} and \cref{eq:A-merged-stabilizers-3}.
All other cases have either disjoint supports or follow by pairwise commutation of elements of \(S_A\) or of the ancilla patch stabilizer generators.
For the first case, fix a pair of stabilizer generators to analyze \((q,k) \in [r_{\bar{A}}]\times [m_A]\).
Expanded out, the pair of stabilizer generators are
\begin{align}
    X_{\iota_{\bar{A}}(k)}X_{\iota_C(k, \ell)} \prod_{j\in [r_{\bar{A}}]} X_{\iota_T(j)}^{H_{\bar{A}}^T[k,j]}&\;&\left(s^Z_{\bar{A},i} |_{A\setminus \bar{A}}\right)Z_{\iota_T(q)} \prod_{i \in [m_A]} Z_{\iota_A (i)}^{H_{\bar{A}}[q, i]}
\end{align}
If \(H_{\bar{A}}[q, k] = 0\) then the stabilizer generators have disjoint support. Otherwise, they must have support on both \(\iota_T(q)\) and \(\iota_{\bar{A}}(k)\).
The other case is proved identically.

We proceed to show that \(\overline{X}_C \overline{X}_A\) is the product of all stabilizer generators defined in \cref{eq:A-merged-stabilizers-T}.
This product is \(\prod_{i \in [m_A]}X_{\iota_{\bar{A}}(i)}X_{\iota_C(i, \ell)} \prod_{j\in [r_{\bar{A}}]} X_{\iota_T(j)}^{H_{\bar{A}}^T[i,j]} = \overline{X}_C \overline{X}_A \left(\prod_{j \in [r_{\bar{A}}]} X_{\iota_T(j)}^{\sum_{i\in [m_A]} H_{\bar{A}}^T[i,j]}\right)\) by \cref{prop:ancilla-patch-params}.
Since the all-ones vector is in the kernel of \(H_{\bar{A}}\), we have that \(\sum_{i\in [m_A]} H_{\bar{A}}^T[i,j]=0\) over \(\F\), completing the proof.
\end{proof}

To prove fault-tolerance of the merging procedure, we use the subsystem formalism~\cite{vuillot2019code, poulsen2017fault} for lattice surgery and map the merging and splitting operation to that of gauge fixing a particular subsystem code.
\begin{definition}[Subsystem code]
  We define a subsystem code on \(n\) qubits by a set of gauge operators \(G \subseteq \mathcal{P}^n\) which generate the gauge group \(\mathcal{G} = \langle G \rangle\).
  We refer to the center of the gauge group \(\mathcal{S}_\mathcal{G} = \mathcal{Z}(\mathcal{G})\) as the stabilizer group of the subsystem code.
  Define the non-trivial dressed logical operators as \(\mathcal{L}_{\textrm{dressed}}=\mathcal{C}(\mathcal{S}_\mathcal{G})\setminus \mathcal{G}\) where \(\mathcal{C}(\mathcal{S}_\mathcal{G})\) is the centralizer of the stabilizer group in the \(n\) qubit Pauli group.
  The dressed distance of a subsystem code is analogous to the distance and is defined to be \(d^{\mathrm{dressed}}:= \min_{L \in \mathcal{L}_{\textrm{dressed}}} |L|\).

\end{definition}
The merging step is done by fixing the middle $X$ checks $\{ S^X_{T,i} \}_{i \in [m_A]}$ (\cref{eq:A-merged-stabilizers-T}) while the splitting step is done by fixing the original boundary $Z$ checks of the A and C codes \(\{s^Z_{\bar{A},i}, s^Z_{C, il}\}_{i \in [r_{\bar{A}}]}\) (\cref{eq:A-merged-stabilizers-2}, \eqref{eq:A-merged-stabilizers-3}).
Importantly, the distance of any complete gauge-fixing of a subsystem code is lower bounded by the dressed distance, and the recovery procedure during the merging/splitting step follows the recovery procedure of the subsystem code.

\begin{theorem}[Subsystem Code Distance]\label{prop:subsystem-distance}
    Let \(S_S\) be the union of the stabilizer generators of a code \(S_A\)  and of a non-degenerate ancilla patch \((S_A, S_B, \overline{X}_A, \overline{Z}_B)\), \(S_M\) be the stabilizer generators of the corresponding \((A,\ell)\)-merged patch, and \(d_A\) (\(d_B\)) the distance of the code defined by \(S_A\) (\(S_B\)).
    Consider the subsystem code defined by the gauge group \(\mathcal{G}\) generated by \(G = S_S \cup S_M \cup \bar{Z}_A\).
    The dressed distance of this subsystem code satisfies \(d^{\mathrm{dressed}} \ge \min\left(d_A, d_B \right) := d_s\).
\end{theorem}
\begin{proof}
  We define the bare logical operators \(\mathcal{L}_{\mathrm{bare}} = \mathcal{C}(\mathcal{G})\setminus{\mathcal{S}_\mathcal{G}}\) such that \( \mathcal{L}_{\mathrm{dressed}} = \{gL | g \in \mathcal{G}, L \in \mathcal{L}_{\mathrm{bare}}\}\).
  We also define the ``split patch'' logical operators \(\mathcal{L}_S = \mathcal{C}(S_S)\setminus \langle S_S \rangle\) where we take the centralizer with respect to the Pauli group supported on the complement of the interface \(T\).
  Let \(\overline{Z}_A\) (\(\overline{X}_B\)) refer to a representative logical operator in \(\mathcal{L}_S\) conjugate to \(\overline{X}_A\) (\(\overline{Z}_B\)).
  Note that $\bar{X}_A \bar{X}_C \notin \mathcal{L}_{bare}$ since $\bar{X}_A \bar{X}_C$ does not commute with $\bar{Z}_A \in \mathcal{G}$ and $\overline{X}_A \overline{X}_C \in \langle S_M\rangle \subset \mathcal{G}$ according to Proposition~\ref{prop:merged-params}.
  Therefore, in terms of the split logical operators, the bare logical operators are 
  \begin{equation}
      \mathcal{L}_{\mathrm{bare}} = \left\{L \in \mathcal{L}_S\setminus \langle \overline{X}_A \overline{X}_C, S_S\rangle \mid L\overline{X}_A \overline{X}_C = \overline{X}_A \overline{X}_C L\right\}.
  \end{equation}
  In particular, this means that \(\mathcal{L}_{\mathrm{bare}}\) contains no elements equivalent to \(\overline{Z}_A\), \(\overline{Z}_C\), or \(\overline{X}_A \overline{X}_C\) with respect to \(\langle S_S\rangle\).
  By \cref{prop:ancilla-patch-params} and the distance of the \(A\)-code, all elements in \(\mathcal{L}_S\) have weight at least \(d_s\).
  
  Since there exists a basis of generators of \(\mathcal{G}\), \(\mathcal{S}_\mathcal{G}\), and of the dressed logical operators that are products of only \(X\) or only \(Z\) (i.e. the code is a CSS subsystem code), it suffices to consider the \(X\) and \(Z\) dressed distances separately.
  We will use the subscript \(X\) or \(Z\) on groups or sets of operators to denote the restriction to only those operators that are products of \(X\) or \(Z\), respectively.

  For the \(Z\) distance, note that \(\mathcal{G}_Z = \langle S_{S,Z}, \{Z_{\iota_T (i)}\}_{i\in [r_{\bar{A}}]}, \overline{Z}_A \rangle\), i.e. the \(Z\) gauge group is generated by the split stabilizer generators, the single qubit \(Z\) operators along the interface, and the $\bar{Z}_A$ logical operator of $A$.
  Fix an operator \(L \in \mathcal{L}_{\mathrm{dressed},Z}\).
  \(L\) has the decomposition \(L = L' g_z \ell\) where \(L' \in \mathcal{L}_{\mathrm{bare},Z}\), \(g_z \in \langle \{Z_{\iota_T (i)}\}_{i\in [r_{\bar{A}}]} \rangle\), and \(\ell \in \langle \overline{Z}_A \rangle \).
  \(\ell L'\) has disjoint support from \(g_z\) and so \(|\ell L' g_z | \ge |\ell L'|\).
  We have that \(\ell \not\simeq L'\), so \(\ell L' \in \mathcal{L}_S\) which implies \(|\ell L'| \ge d_s\).
  Overall, we conclude \(|L| = |\ell L' g_z | \ge |\ell L'| \ge d_s\).

  For the \(X\) distance, we will use a ``topological" argument and show that the support of any logical operator cannot be completely eliminated from certain sets.
  
  First note that \(\mathcal{G}_X = \langle S_{S,X}, G_T \rangle\) where \(G_T=\{S^X_{T,i}\}_{i\in [r_{\bar{A}}]}\) is as defined in \cref{eq:A-merged-stabilizers-T}.
  Let \(i \in [m_A]\) \(R_i = \iota_{VV}(i, [m_B]) \cup \iota_{\bar{A}}(i)\) be the \(i\)-th row of the \(VV\) qubits of \(C\) along with the element of \(\bar{A}\) adjacent to it.
  Since \((1,1,\dots,1)^T \in \ker H_{\bar{B}}\), any element of \(\langle S_{C,X}\rangle\) or \(\langle G_T\rangle \) must have an even support in each row \(R_i\).
  We will use \(| \cdot |_\mathrm{row}\) to denote the number of rows \(\{R_i\}_{i \in [m_A]}\) in which an operator has support on an odd number of qubits.
  This satisfies the property that for any two operators \(\alpha,\beta \in \{I,X\}^{\times A \cup C \cup T}\), \(|\beta|_\mathrm{row} = 0\) (e.g. elements of \(S_S\) and \(G_T\)) implies \(|\alpha\beta|_\mathrm{row} = |\alpha|_\mathrm{row}\).

  Fix an operator \(L \in \mathcal{L}_{\mathrm{dressed},X}\).
  \(L\) has the decomposition \(L = g L'\) where \(L' \in \mathcal{L}_\mathrm{bare}\) and \(g \in \langle G_T \rangle\).
  \(L'\) has no support on \(T\), so \(L'\) has the further decomposition \(L' = L_A L_C\) where \(L_A, L_C \in \mathcal{C}(S_S)\), \(L_A\) (\(L_C\)) is supported only on \(A\) (\(C\)), and at least one of \(L_A\) or \(L_C\) is in \(\mathcal{L}_\mathrm{bare}\).
  
  In the following, we will use \((\cdot )_S\) to denote the restriction \(\cdot |_S\).
  Since \(\overline{X}_A \overline{X}_C\) is in the gauge group, there exists \(g' \in \langle S_{C,X}, G_T\rangle\) such that \(L_C = L_C' g'\) where \(\supp L_C' \subseteq \bar{A}\) i.e. \(L_C' \in \{\overline{X}_A, I\}\).
  Then,
  \begin{align}
      |L | &= |L_A L_Cg |\\
      &\ge \left|\left(L_A L_C' g' g\right)_{A \setminus \bar{A}}\right| + \left|\left(L_A L_C' g' g \right)_{\bar{A} \cup C} \right|\\
      &\ge \left|\left(L_A L_C' g' g\right)_{A \setminus \bar{A}}\right| + \left|L_A L_C' g'g\right|_\mathrm{row} \\
      &= \left|\left(L_A L_C'\right)_{A \setminus \bar{A}}\right| + \left|L_A L_C'\right|_\mathrm{row} \\
      &= \left|\left(L_A L_C\right)_{A \setminus \bar{A}}\right| + \left|\left(L_A L_C'\right)_{\bar{A}}\right| \\
      &= \left|L_A L_C'\right| \\
      &\ge d_s
  \end{align}
  Where we have used that \(|g'|_\mathrm{row}=0\) and that \(\supp g' \cap (A \setminus \bar{A}) = \emptyset\).
  Since \(L_A \not\simeq L_C'\), \(L_A L_C' \in \mathcal{L}_\mathrm{bare}\), so that \(|L_A L_C'| \ge d_s\).
\end{proof}

The case of the \(B\)-merge proceeds by a nearly identical argument with appropriate substitutions of \(A \leftrightarrow B\) and \(X \leftrightarrow Z\).
Note that we do not analyze the case of measurement errors.
By repeating measurements a number of times proportional to the distance, we believe that an adaption of the argument in   Ref.~\cite{gottesman2013fault} is sufficient to prove fault tolerance against circuit level depolarizing noise.

\subsection{Parallel teleportation}
In the previous section, we analyzed the fault tolerance of joint logical measurements required for teleportation between a single logical qubit in a qLDPC block and a surface code.
This scheme straightforwardly extends to teleportation of certain qLDPC logical qubits in parallel using a separate ancilla patch for each logical qubit and its paired surface code and merging the qLDPC/surface blocks with all the ancilla patches simultaneously. 
For a HGP code, we can find multiple logical qubits for which there exists logical representatives satisfying the requirements of  \cref{prop:ancilla-patch-params} and \cref{prop:subsystem-distance}. Thus, each formed ancilla patch has a distance lower bounded by $d_s$, the minimum of the HGP distance and the surface code distance.
The dressed distance of the subsystem code describing the merged code can also be shown to be at least $d_s$ by inductively applying an argument essentially identical to the proof of \cref{prop:subsystem-distance} with slight modifications due to the slightly expanded gauge group.
Since the lattice surgery is done to disjoint subsets of qubits and incident checks, the code remains LDPC.

Let \(k\) be the desired number of logical qubits of the HGP code.
Fix a $[n_0, k_0 = \sqrt{k}, \Theta(\sqrt{k})]$ classical linear code defined by the check matrix $H$ satisfying $\ker H^T = 0$, and consider the hypergraph product code \((H,H)\).
Theorem 1 of \cite{quintavalle2022partitioning} furnishes a basis of logical representatives for the $k$ inequivalent logical $X$ operators supported on $k_0$ distinct columns of the \(VV\) qubits.
Label the basis by coordinates \([k_0] \times [k_0]\) where the logical operator \((i,j)\) has support only in the \(j\)-th column, \(\iota_{VV}([n_0], j)\).
Then, we may simultaneously teleport any set of \(m \le k_0\) logical qubits with distinct column coordinates to \(m\) separate surface code patches.
By taking \(m = O(\sqrt{k} / \mathrm{poly} \log kT)\), the total ancilla size is $O(m \sqrt{k} d_{\mathrm{comp}}) = O(k)$, where the computation codes distance $d_{\mathrm{comp}} = O(\mathrm{poly}\log kT)$ is chosen such that it suffices for suppressing the logical failure probability of a depth-$T$ logical circuit. As a result, we can maintain constant space overhead during logical computations.